\documentclass[prodmode,acmtaas]{acmsmall} 

\usepackage[ruled]{algorithm2e}

\SetAlFnt{\small}
\SetAlCapFnt{\small}
\SetAlCapNameFnt{\small}
\SetAlCapHSkip{0pt}
\IncMargin{-\parindent}

\usepackage{amsmath}
\usepackage{hyperref}
\usepackage{soul}

\usepackage[lofdepth,lotdepth]{subfig}

\usepackage{algorithmic}
\usepackage{array}
\usepackage{url}
\usepackage{amssymb}
\usepackage{color,soul}
\usepackage{lineno,xcolor}

\usepackage{fancyhdr}
\begin{document}

\markboth{Semwal et al.}{Ordering Multi-Robot Task Executions within a CPS}

\title{On Ordering Multi-Robot Task Executions within a Cyber Physical System}
\author{TUSHAR SEMWAL
	\affil{Indian Institute of Technology Guwahati}
	SHASHI SHEKHAR JHA
	\affil{Indian Institute of Technology Guwahati}
	SHIVASHANKAR B. NAIR
	\affil{Indian Institute of Technology Guwahati}}

\begin{abstract}

With robots entering the world of Cyber Physical Systems (CPS), ordering the execution of allocated tasks during run-time becomes crucial. This is so because, in a real world, there can be several physical tasks that use shared resources that need to be executed concurrently. In this paper, we propose a mechanism to solve this issue of ordering task executions within a CPS which inherently handles mutual exclusion. The mechanism caters to a decentralized and distributed CPS comprising nodes such as computers, robots and sensor nodes, and uses mobile software agents that knit through them to aid the execution of the various tasks while also ensuring mutual exclusion of shared resources. The computations, communications and control, are achieved through these mobile agents. Physical execution of the tasks is performed by the robots in an asynchronous and pipelined manner without the use of a clock. The mechanism also features addition and deletion of tasks and insertion and removal of robots facilitating \textit{On-The-Fly Programming}. As an application, a Warehouse Management System as a CPS has been implemented. The paper concludes with the results and discussions on using the mechanism in both emulated and real world environments.

\end{abstract}

\ccsdesc[500]{Computing methodologies~Intelligent agents}
\ccsdesc[300]{Computing methodologies~Multi-agent systems}
\ccsdesc[300]{Computing methodologies~Mobile agents}
\ccsdesc[300]{Computing methodologies~Cooperation and coordination}
\ccsdesc[300]{Computing methodologies~Distributed algorithms}
\ccsdesc[300]{Computer systems organization~Embedded and cyber-physical systems}
\ccsdesc[300]{Computer systems organization~Robotics}
%
%


\keywords{Mutual Exclusion, Distributed  Systems, Multi-Robot Systems (MRS).}


\begin{bottomstuff}
Primary Author's address: Tushar Semwal (t.semwal@iitg.ernet.in), Department of Computer Science and Engineering, Indian Institute of Technology Guwahati;
(Current address) Department of Computer Science and Engineering, Guwahati, Assam 781039.
\end{bottomstuff}

\maketitle

\section{Introduction}

While robotic applications are fast making inroads into a plethora of automated systems, the tight coupling between the application and the robotic hardware seem to deter both their scalability and flexibility. The need of the day is to transform such automated systems into ones that are malleable and accessible over a network. Through this transformation, a fair amount of generic nature can be embedded within such systems, thereby allowing for changes to be made in the patterns or nature of executions of the tasks performed. This flexibility  can be realized only if we facilitate networking among all the entities within these systems. Networking can allow the entities to communicate with one another and resolve several issues that crop up during run time. If the entities are mobile, the network becomes dynamic and makes one-to-one communication, a much disorganized task. A centralized approach for controlling the entities may perform well but makes the system rigid, expensive and hardly scalable. On the contrary, a decentralized and distributed control mechanism coupled with a mobile computing environment can empower these systems with autonomy, flexibility and scalability. Such automated scenarios can be viewed to be made up of two basic components -- a \textit{cyber} component that caters to both computing and networking of  the entities and a set of \textit{physical} processes which are executed in the real world by a set of robots using percepts received from either on-board sensors or sensor nodes. Considering the fact that the physical processes are initiated, linked and to some extent controlled by the cyber component, this type of a system can be categorized as a Cyber-Physical System (CPS) \cite{Baheti2011a}. Hence, a networked Multi-Robot System (MRS) coupled with a mobile computing environment can provide a fitting framework for a CPS.

\textit{\textbf{Research in MRS}} has mostly been focused broadly on two main areas viz. \textit{task allocation} and \textit{task partitioning}. In the former \cite{Gerkey2001}, tasks are assigned to the appropriate participating entities (robots) in such a way that a desired performance level can be achieved with complete utilization of available resources. The latter, on the other hand, is the process by which a task is divided into a set of subtasks so as to reduce the complexity of its execution \cite{Ratnieks1999}. Apart from these, there is also a third objective crucial to an MRS based CPS viz. that of \textit{task execution} which is grossly ignored in MRS specifications. Task execution is an inherent objective (usually defined by the user) that always commences after task allocation or partitioning. While the allocation and partitioning are merely planning models, task execution adheres to the actual implementation which validates the assignments of the tasks. Hence, both task allocation and task partitioning are dependent on task execution without which a task cannot be said completed. Early work on Multi-Robot Task Allocation (MRTA) by Parker \citeyear{Parker1998} describes an architecture where fault tolerance was incorporated in a  heterogeneous set of robots for carrying out different tasks. They demonstrated dynamic task allocation (a subclass of task allocation) within an MRS. A formal analysis of the problems faced in MRTA has been presented in \cite{Gerkey2003}. Botelho and Alami \citeyear{Botelho1999} describe a technique for allocation and reallocation of tasks. In their work, each robot is provided with details of its own plan. A robot is allowed to make changes in its plan depending upon its capabilities as also those of the other robots. The use of auctioning techniques based on dynamics of a market has been proposed by Dias and Stentz \citeyear{dias2000free}, where the robots are assigned tasks through negotiations with their peers in a distributed manner. Khaluf and Rammig \citeyear{Khaluf2013} narrow down the scope of task allocation to time-constrained tasks. However, they have ignored the complexities of real-world task execution and provided only the simulation results.

The approaches discussed so far do not address the complexities involved in ordering of actual task executions in the real world. Ordering such executions requires a careful understanding of the available resources needed to complete a task within a CPS of robots. These resources may be either exclusively available to every robot or they may need to be shared. If the resources are to be shared by the robots within a CPS for the completion of their assigned tasks, then a mechanism for mutual exclusion of shared resources becomes mandatory in order to avoid contentions and deadlocks. Tasks involving shared resources are common in the real-world. For instance, a ticket counter where people wait in a queue, is a typical example of a shared resource. In the realm of MRS, a sole battery-charging terminal where at an instant only one robot can plug-in and charge itself, forms an example where mutual exclusion needs to be exercised. The mutual exclusion problem in an MRS can be solved using centralized control. The central controller can monitor and communicate with all the robots regarding their turns to gain access to all the shared resources. Though simple and straightforward, this solution could drastically load the central controller with heavy computational and communication overheads. Further, any change in the CPS would mean bringing down the central server.

In the domain of distributed computing, the \textbf{\textit{Mutual Exclusion of shared Resources} (MER)} is referred as a classical benchmark problem to resolve resource contentions \cite{raynal1986algorithms}. MER is required when different nodes need to access a shared resource at the same time, lest a race condition \cite{raynal1986algorithms} occur. 
A CPS constituting mobile networked robots could be looked upon as a Mobile Ad-hoc Network (MANET). The problem of MER becomes more complicated in the context of MANETs \cite{giordano2002mobile} wherein mobile nodes move in a disorganized manner leading to dynamism in the communication topology. In addition, MANETs are constrained with limited bandwidth, low power usage, low computations capabilities, dynamic topology, etc. \cite{Fife:2003:RID:776985.776991} that increases the complexity of the MER problem as compared to their static counterparts. Solutions to the MER problem in distributed and dynamic networks can be broadly divided into two categories \cite{saini} \textemdash\ \textit{token} based and \textit{permission} based. In the token based approach, a node with a unique token can access the shared resources while others have to wait for the arrival of the token. On the other hand, in the permission based approach, a node can get access to a shared resource if it can get permissions from all other nodes in the network by exchanging messages.  Since, in this approach, a node sends a request for getting access through messages to all the connected nodes, it consumes bandwidth and thereby introduces network latency. Although many variants of the MER problem have been proposed \cite{Chandy:1984:DPP:1780.1804,groupmutual:2001:NGM:383962.383997,kmutex,localmutex}, an adaptive and scalable solution in context of a CPS, wherein the entities performing tasks need to share physical resources in the real world, has still not been proposed. Wu et al. \citeyear{wuweigang}, have modeled the problem of mutual exclusion of traffic intersections as a variant of the classical mutex problem. Vehicles compete to get access to the traffic intersection by exchanging messages. A vehicle passes through the intersection when it receives permissions from other vehicles involved in the competition. Their approach however, uses multiple messages which lead to communication overheads and network latency. Minimizing such overheads while ensuring MER is crucial for the performance of a CPS.

Depending upon their nature, tasks can be divided into two types - (i) Independent and (ii) Sequential. Independent tasks can be executed in isolation and thus do not in any way rely on other tasks. By sequential, we mean that these tasks follow a topological order such that a task say, $T_{i+1}$ is executed only if the execution of the preceding task $T_{i}$ is completed. In a typical computing environment, when a program includes both parallel and sequentially executable instructions or methods, the associated compiler separates the independent ones from the sequential ones. Based on the program, it assigns the independent ones to individual cores within a multi-core processor to maximize parallelism. The number of such cores which could be looked upon as independent processing units, naturally do not change.
On the contrary in a real world multi-mobile-robot scenario where robots are synonymous to such computing cores, this may not be the case. The number of robots available to perform a set of tasks may vary over time. Such variation could be due to the fact that some robots may need to be charged while others could have malfunctioned for some reason. Their number could also increase, if more robots are deployed into the scenario. A precompiling procedure to initially allocate sequential and independent tasks to a set of robots, as in a typical multi-core computing environment could be disastrous. 

Further, in the physical world, tasks could be \textit{interdependent} by virtue of the fact that they require both robots and resources to get executed. For instance, consider the case where robot $R_1$ is to execute a task $T_1$ using a resource $\psi_1$ while robot $R_2$ is to execute task $T_2$ using the same resource $\psi_1$. In this scenario, assuming $T_1$ and $T_2$ to be independent tasks, it can be observed that though both robots $R_1$ and $R_2$ are free to execute the two tasks, the non-availability of $\psi_1$ concurrently to both $R_1$ and $R_2$ creates a bottleneck. One of them has to wait for the other to free the resource $\psi_1$ forcing  $T_1$ and $T_2$ to be executed sequentially executed. 
It may be noted here that, independent tasks may also suffer from similar bottlenecks when they require the same resource. Under such conditions, this \textit{resource dependency} forces these independent tasks to be executed sequentially.
One can thus conclude that a technique that can handle the ordering of all types of tasks \textit{on-the-fly} while also catering and effectively utilizing the varying number of mobile processing units, forms a \textit{sine qua non} for CPSs comprising mobile robots.

\textit{In this paper}, we formulate the problem of ordering the execution of sequential, independent and interdependent tasks to be executed by multiple mobile robots within a CPS and propose a mechanism to solve the same. An agent based approach has been formulated to ensure MER among multiple robots connected to form a dynamic network. A sequence of topologically ordered and interdependent tasks that involves shared resources, forces their execution in the form of a pipeline. Since the number of mobile robots available to execute a set of tasks could vary, we have tried to portray these robots as a pipeline wherein the number of processing units could vary during run-time. A conventional pipelined computing architecture requires a clock in order to synchronize and allocate proper time slots for the execution of processes. The execution times for the various tasks performed by a set of robots however, vary over time due to several real-world problems.
If a pipeline needs to cater to such varying times required for the executions, it should possess an inherently adaptive clocking mechanism so as to compensate for such variations. 

The algorithm proposed in this paper is novel in the sense that: 1) In this work, we have used intelligent messages in the form of mobile agents to solve the problem of mutual exclusion while executing tasks in a multi-robot distributed environment. Conventional distributed scenarios as in Wu et al. \citeyear{wuweigang}, use message broadcasts to share information and ensure mutual exclusion of shared resources.  Message broadcasting drastically increases the communication cost \cite{wuweigang} and can clutter a network. In the proposed mechanism, we have used a conscientious agent migration strategy \cite{minar1999} which has least inter-node communication cost \cite{godfrey2013mobileAgntIoT} as compared to other agent based approaches such as CLInG \cite{sempe_evap}, EVAP \cite{chu_CLInG} and Random-walk with cloning \cite{Gaber2008}.
2) Synchronization in distributed settings is a major challenge and is traditionally achieved by using a single node (or a subset of nodes) which provide for clocking. This poses issues of reliability when such nodes fail. In the domain of robotics, the problem of synchronization deteriorates since the time required to execute a given task by a robot can vary due to several environmental factors. In the mechanism proposed herein, the agents ensure an inherent adaptive clocking mechanism to achieve synchronization across the network of robots. In addition, features such as concurrent execution of tasks, \textit{on-the-fly} addition and deletion of tasks and inclusion and removal of robots, emphasize the flexibility and versatility of the proposed mechanism. 
Finally, a Warehouse Management System (WMS) as an application has been implemented to demonstrate the feasibility of our approach.

In brief, our major contributions towards the Task Execution Ordering Problem (TEOP) are - 
\begin{enumerate}
	\item A mobile agent based distributed mechanism for ordering multi-robot task executions.
	\item A solution for the MER problem among multiple robots within a CPS. 
	\item Validation of the proposed mechanism through emulation.
	\item Real world implementation of the proposed mechanism with WMS as an application.
\end{enumerate}

The remaining part of paper is organized as follows: Section 2 discusses agent based
systems and their applications. Section 3 describes the constituents and system specifications
of the proposed CPS. The Task Execution Ordering Problem (TEOP) among multiple robots and the
inherent objectives for realizing the CPS are discussed in Section 4 while the proposed
mechanism is described in Section 5. Section 6 describes a real-world implementation
of the proposed mechanism while Section 7 presents the results obtained in both the
emulated and real-world scenarios. Finally, Section 8 concludes the paper and provides
directions for future work.

\section{Agent based Systems}
An agent is as an intelligent software code that has a certain degree of autonomy \cite{Franklin}. Agents are smart beings that reside in the cyber world and carry out tasks or computations on behalf of the users. As human beings in the real-world, agents form their counterpart in the cyber world. They are autonomous, decisive, flexible, adaptive, reactive, pro-active, social, have locality of reference and many other features \cite{schumacher2001multi}. Agents can be broadly divided into two types \textemdash\ static agents and mobile agents. Usually agents are stationary entities which occupy a fixed location within a networked environment. Mobile agents on the contrary are distinguished by their mobility which allow them to move freely within a network of nodes. Since, mobile agents form an important component of the proposed mechanism for executing and ordering of a sequence of interdependent tasks within a CPS of multiple robots, the succeeding section presents a brief background on the use of these agents.

\subsection{Mobile Agents}
As mentioned, a mobile agent is a program that can migrate from one node to another within a network and can perform autonomous computations. Along with mobility, these agents also possess other abilities such as cloning, autonomy, payload carrying capability, on-node execution, local decision making, adaptability, etc. \cite{outtagarts2009mobile}. Mobile agents have been used in a myriad of applications ranging from wireless sensor networks \cite{Chen2007}, e-commerce \cite{Maes1999}, robot control \cite{Kambayashi,Godfrey2008}, security \cite{Boukerche2007} and e-learning \cite{Zaiane2002}. Posadas et al. \citeyear{Posadas2008} highlight the advantages of using mobile agents for controlling mobile robots using two approaches of executing tasks. In the first approach, all the mobile robots communicate with a central server to get the necessary actions which aid in the completion of a set of tasks. In the second, the mobile agents are released into the network of mobile robots. These are then made to execute programs locally on each mobile robot based on their local decision making capabilities. The authors report that the latter approach is more effective in terms of  time required to execute an action by a mobile robot when compared to the former centralized method. This is so because considerable time is wasted in communication with the central server. 

Godfrey and Nair \citeyear{Godfrey2012} describe how mobile agents can be used to provide services in an MRS. They also compare the performance of mobile agent migration strategies. Some of the important reasons for the use of mobile agents for the realization of distributed systems are \cite{cruz2011handbook} \textemdash
\begin{enumerate}
	\item  \textit{Network traffic and latency reduction}: Mobile agents perform computations and interactions locally at a node. This results in faster response and avoids excessive message passing.
	\item \textit{Adaptation and customization}: In traditional distributed systems, as the protocols evolve for transmitting and interpreting the outgoing and incoming data respectively, it becomes cumbersome to update the servers and creates a legacy problem \cite{Lange1998}. In such scenarios, mobile agents provide for flexibility as clients can dispatch them to the server for establishing the amendments to the protocols.
	\item \textit{Robustness and fault tolerance}: In dynamic and distributed networks, it is impractical to continuously maintain static communication links among the nodes. In such cases, the tasks required to be executed can be embedded within the mobile agents. These agents can then be dispatched into such dynamic networks. After being dispatched, these agents become independent and operate autonomously by carrying out executions at the designated nodes in the network. The results can be accumulated later by reconnecting with these agents.
\end{enumerate}

Mobile agents thus can serve as an effective tool for realizing distributed mechanisms over a network of nodes. 
\subsection{Mobile Agent Framework}
A mobile agent framework is an execution environment or a platform which provides support for agent development, programming and deployment within a network. It provides tools for the users to create and manage agents and their behaviours. The agents (static and  mobile) are managed by the framework in order to ensure their successful execution and operation. There are various mobile agent frameworks available in a variety of programming languages such as Java, C/C++ or Prolog. JADE \cite{Bellifemine2001}, a Java based agent framework, is known for its simple development process along with FIPA 2000 compliance. Mobile-C \cite{Mostinckx2009} is a purely C/C++ based agent framework which, due to its small code size, readily supports embedded devices. JINNI \cite{tarau1999jinni}, Typhon \cite{Matani2011} and \textit{Tartarus} \cite{Semwal2015,semwal2016tartarus}, being Prolog based frameworks, facilitate rapid prototype development. The work reported herein uses \textit{Tartarus}, a mobile agent framework for the  development and management of static and mobile agents. We have chosen \textit{Tartarus} as it supports rapid prototyping, multi-threaded execution and faster mobility. In addition, it comes with a pre-installed plug-in channel which allows for the control and interaction with other embedded devices and sensors.

\section{Preliminaries}
This section presents the entities and characteristics that make up the CPS used herein followed by a formal description of the problem at hand. Further, we discuss the mechanisms to ensure MER when the tasks within a CPS need to be executed in a sequential manner. The manner in which the tasks can be altered, added or removed \textit{on-the-fly} in/from the sequence is also be illustrated.

\subsection{Constituents of the proposed CPS}
A CPS is an amalgam of both the cyber and the physical worlds where the term \textit{cyber} comprises computations, communications and control while the term \textit{physical} comprises interactions with the real world \cite{Shi2011}. Our proposed CPS is composed of heterogeneous entities such as a Multi-Robot System (MRS), mobile agents, sensors and computer nodes. Mobile agents form the cyber entities which carry out computations, manages all the communications and control the dynamics of the MRS. The interaction of robots with the external surroundings where robots execute the sequential tasks forms the physical component. Following are the basic constituents of the proposed CPS under consideration:
\begin{enumerate}
	\item Nodes:  A node refers to any device that is capable of computations and communications  and hosts an agent framework. It can be an embedded system, a personal computer, a robot or even a sensor node. Nodes are connected to each other to form a network \textbf{\textit{W}}.
	\item Network: A network \textbf{\textit{W}} is a dynamic wireless Mobile Ad-hoc Network (MANET) wherein a node can connect or disconnect to another node at any point of time. The connections are inherently managed by the nodes within the network using any of the available mechanisms \cite{Dr.S.S.Dhenakaran2013}.
	\item Robots: A set of networked robots \textit{\textbf{R}} =\{$R_{1}, R_{2}, R_{3},\dots, R_{k}$\textbar $k\geq1$\} all of which hosts an  agent platform within and can connect to the network \textit{\textbf{W}} in an ad-hoc manner. These are essentially a subset of nodes responsible for the execution of tasks. Robots are mobile and are equipped with sensors and actuators that enable them to sense their environment and act upon them, respectively.
	
	\item Tasks: A set of finite tasks \textit{\textbf{T}} = \{$T_{1}, T_{2}, T_{3},\dots, T_{n}$\textbar $n\geq1$\}, capable of being executed by the set of robots \textbf{\textit{R}}. 
	 
	\item Resources: Utilities and nodes other than robots, such as a path, parking/charging bays, a  rack containing items which can act as a node, sensor nodes, etc.,  in the MRS environment required by a robot to accomplish a task, constitute a set of resources $\boldsymbol\psi$ = \{\textit{$\psi$$_{1}$}, \textit{$\psi$$_{2}$}, \textit{$\psi$$_{3}$},\dots,\textit{$\psi$$_{r}$}\textbar $r\geq1$\}. Resources need to be shared amongst robots in the set \textit{\textbf{R}} while a robot executes the tasks in \textit{\textbf{T}}. Once a robot takes over a resource(s), it becomes non-shareable before it is freed by the robot. For clarity, two forms of conventions have been followed in this paper viz. $\psi_i$ and $\psi^i$, where for the task $T_i$, $\psi_i$ is a particular resource from the set $\boldsymbol{\psi}$ while $\psi^i$ $\subseteq \boldsymbol\psi$.
	
	\item States:  States pertain to robots. $S_i^{T_j}$ indicates that a robot  is in state $S_i$ and requires to execute the task $T_j$. 
	All free robots remain in the state designated as $S_1^{*}$.
	
	\item Agents: A set of mobile agents $\boldsymbol\mu$= \{\textit{$\mu$$_{1}$}, \textit{$\mu$$_{2}$}, \textit{$\mu$$_{2}$},\dots,\textit{$\mu$$_{m}$}\textbar $m\geq1$\}, such that each mobile agent $\mu_{i}\in\boldsymbol\mu$, carries the programs  of its associated tasks as its payload. It may be noted that each agent carries the programs for a set of task(s) assigned to it along with the information of the required set of associated resources. An agent also carries with it the State Information (SI) in the form of  $S_i^{T_j}$ of the robots which it can serve, and the next state to which the robots transit after execution of $T_j$. 
	\item Job: A Job $J_i$ is a collection of tasks in  \textit{\textbf{T}} along with the associated set of resources in $\psi$, which are required to be executed by the robots in \textit{\textbf{R}} and constitute the basic inputs to the system. Here, $J_i$ $\subseteq$ \{($T_{1}^i$,$\psi^1$),($T_{2}^i$,$\psi^2$),..,($T_{n}^i$,$\psi^n$)\}, $T_n^i$ is the $n^{th}$ task of job $J_i$ and $\psi^n$  $\subseteq \psi$. The intersection of subsets of the type $\psi^n$ need not be a null set indicating that a particular set of resources could be required by more than one task.  These jobs are processed and packed into mobile agents by a Job Distributor $J_{Dist}$. New jobs received by the $J_{Dist}$ could commence their execution even when their predecessors are being executed.
\end{enumerate}
Here \textit{k}, \textit{n}, \textit{r} and \textit{m} $\in$ I where I is a set of positive integers.

\subsection{System Specifications}
For a better insight into the complexity of the proposed CPS, the specifications and behavior of the system need to be defined precisely. Listed below are some pertinent points about the system \textemdash
\begin{enumerate}
	\item The number of nodes in the network \textit{\textbf{W}} is finite.
	\item Any node can connect or disconnect from the network \textit{\textbf{W}} at any instant of time.
	\item The system is completely oblivious of the total number of robots \textit{\textbf{R}} present in the network \textit{\textbf{W}} at any point of time.
	\item The number of mobile agents $\boldsymbol{\mu}$ inhabiting the network \textit{\textbf{W}} varies dynamically with the change in the sequence of tasks.
	\item The sequence in which the tasks in the set \textit{\textbf{T}} need to be executed may be changed as per the requirements.
	\item Each of the robots and the agents are autonomous entities capable of carrying out independent executions. 
	\item There is no direct robot-to-robot or agent-to-agent communication.
\end{enumerate}

\section{The Task Execution Ordering Problem (TEOP)}
Consider a CPS with a finite number of homogeneous robots. Each robot is required to carry out the execution of a finite number of tasks that are interdependent. Since all the robots are required to execute such tasks, a robot may need a set of resources which are shared among its peers. This invokes the necessity for the mutual exclusion of these resources while executing the tasks.  As discussed earlier, in the real world the number of robots available for task execution may vary with time. In addition, one may need to alter, add or delete tasks \textit{on-the-fly}. Under such circumstances, ordering the task executions \textit{on-the-fly},  becomes a challenging problem.
We have modeled this problem as a Task Execution Ordering Problem (TEOP) and proposed a solution to the same using a set of mobile agents. 

\begin{figure}[t]
	\centering
	\centerline{\includegraphics[width=120mm]{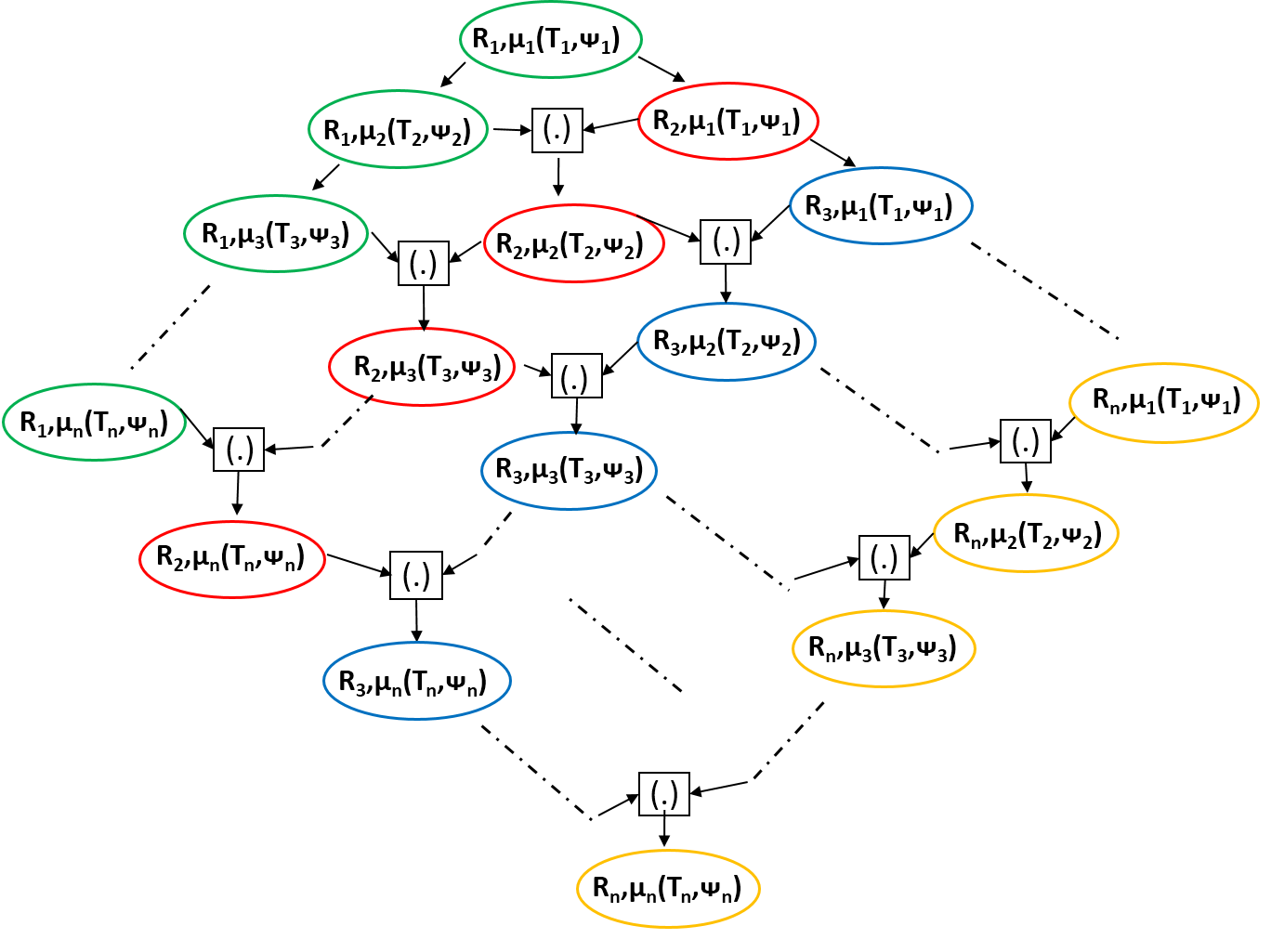}}
	\caption{Graph depicting the inherent sequential and interdependent nature of execution of tasks}
	\label{fig:one}
\end{figure}

For simplicity, consider a straightforward scenario where each task $T_i$ requires a single resource $\psi_i$. The scenario can be easily extended to more complex ones where instead of a single resource, a task may require a set of resource $\psi^i$ which may be common with those of other tasks. Figure~\ref{fig:one} represents the problem in the form of a directed acyclic graph. Each node of the graph represents a robot $R_{i}$ performing a task $T_{i}$ using the resources in $\psi$$_{i}$ with the help of a mobile agent \textit{$\mu$$_{i}$}. Additionally, there are some nodes which represent the operator - AND (.). This operator makes sure that a certain robot can perform a task $T_{i}$ if and only if the previous dependent task $T_{i-1}$ is completed (the sequential execution constraint). Since mobile agents carry the programs of the corresponding tasks, this translates the actual dependency of execution onto these agents. Hence, if a robot $R_{j}$ is vying for resources to execute a task $T_{i}$, the mobile agent $\mu$$_{i}$ carrying the program for the task $T_{i}$ must be free and available in the network. By free, we mean that the mobile agent should not be resident within a robot nor aiding the execution of the associated task.

The graph shown in the Figure~\ref{fig:one} depicts the manner in which the robots execute the tasks while also ensuring mutual exclusion. It may be noted that all the nodes having the same colour correspond to the same robot. Thus only one of these same-coloured nodes can be active at any moment of time. For instance, the red coloured nodes stand for the robot $R_{2}$. The tasks $T_{1}$, $T_{2}$ and $T_{3}$ thus cannot be executed concurrently since they all need to be executed by the same robot viz. $R_{2}$.

As an explanatory example, consider the node $(R_2, \mu_2(T_2,\psi_2))$ denoting execution of task $T_2$ by robot $R_2$ using resource $\psi_2$ and mobile agent $\mu_2$. In order to traverse to this node, both inputs to the AND node viz. $(R_1,\mu_2(T_2,\psi_2))$ and $(R_2,\mu_1(T_1,\psi_1))$ need to be TRUE i.e. $R_1$ should have executed task $T_2$ using $\mu_2$ and $\psi_2$ and $R_2$ should have executed $T_1$ using $\mu_1$ and $\psi_1$. This indicates the sequential nature of execution of tasks by $R_2$ viz. $(R_2,\mu_1(T_1,\psi_1)) \rightarrow (R_2,\mu_2(T_2,\psi_2))$. Additionally, the interdependency between $(R_2,\mu_2(T_2,\psi_2))$ and its predecessor nodes $(R_1,\mu_2(T_2,\psi_2))$ and $(R_2,\mu_1(T_1,\psi_1))$ can also be observed. This means that $T_2$ (carried only by agent $\mu_2$), which requires resource $\psi_2$ for execution, cannot be executed by multiple robots at the same time thereby ensuring mutual exclusion.

\subsection{The Inherent Objectives}

With several robots and shared resources, ordering the executions of sequential, independent and interdependent tasks, becomes a complex task especially when the number of executing robots and tasks vary at run-time.  This section discusses this problem of ordering in terms of its segregated objectives.

\textbf{Objective 1.}
The main objective of the work presented in this paper is to honour the mutual exclusion of the use of resources $\boldsymbol\psi$ by the robots in \textbf{\textit{R}} while executing all the tasks in the set \textit{\textbf{T}}. Let $R_{i}^{j,t}$ be a binary function that returns $1$  during the time slot when the robot $R_{i}$ has acquired resource $\psi_j$. Hence the objective is
\[\ \forall R_i\in\textit{\textbf{R}}, execute(R_{i}, \textit{\textbf{T}}) \]
subject to \begin{equation}
	\forall \psi_{j}\in\psi, \sum ^{n}_{i=1} R_{i}^{j,t} \leq 1
	\label{eqn:one}
\end{equation}

\[ R_{i}^{j,t} \in \{0,1\}, \forall i, j \] at any time instant $t$. \\
Here,
\textit{execute}($R_{i}$, \textit{\textbf{T}}) denotes that the robot $R_{i}$ $\in$ \textbf{\textit{R}} executes the tasks in the set \textit{\textbf{T}}. 
As can be observed from Equation \ref{eqn:one}, the constraint of the objective function essentially denotes the MER amongst the robots such that no more than one robot can acquire the same resource at any given time.

The Objective 1 essentially makes the robots in \textbf{\textit{R}} to align their executions in the form of a \textit{pipeline}. Pipelining \cite{null2014essentials} is extensively used by the computer processors in order to increase throughput. It facilitates the execution of the several of instructions in a single unit of time. For instance, the three main subtasks performed by processors to complete the execution of an instruction are \textendash \ \textit{Fetch}, \textit{Decode} and \textit{Execute} \cite{null2014essentials}. In the absence of a pipeline, the processor has to finish the first instruction which it received from the memory and then move towards the next instruction sequentially. This makes the other functional units of the processor such as the ALU to idle while the \textit{Fetch} instruction is being performed. However, in a pipelined architecture, when the processor is busy executing an instruction, other units within, can perform other subtasks concurrently. However, these subtasks need to be synchronized by a common clock. Any increase or decrease (addition/deletion) in the number of subtasks can cause asynchronism. This gives rise to our second objective.

\noindent\textbf{Objective 2.}
The second objective is concerned with maintaining the time period of each stage in the asynchronous robotic pipeline. A pipeline in the context of processors comprises a set of cascaded tightly coupled processing elements. The output of one is given as input to the next. These elements are driven \textit{synchronously} by a clock whose time period is set to a value greater than the maximum delay incurred between the elements in the pipeline. Finding this maximum time delay and setting the clock accordingly is possible in the domain of a computing system as the execution and delay times once fixed, never change. However, in real world robotic scenarios, these timings depend on the task at hand and the conditions and position or location of the robot. In other words, the time to execute a task could vary temporally. This adds another dimension of complexity since the \textit{asynchronously} executed tasks whose execution times vary, could cause problems when the robot(s) try to access a shared resource. Under such conditions the use of a synchronous lock whose time period is set to a constant value \textit{a priori} could prove to be disastrous. It may be noted that when \textit{n} robots are executing \textit{n} tasks, each  with distinct resources concurrently, the robotic pipeline is full and operating at its maximum, thus satisfying the following optimality criteria \textendash
\begin{equation}
	\forall\psi_j\in\psi, 
	\sum\limits_{j=1}^{r}\sum\limits_{i=1}^{k} R_i^{j,t} = r
	\label{eqn:three}
\end{equation} 
at any time instant $t$.\\

\begin{figure}[t]
	\centering
	\centerline{\includegraphics[width=130mm]{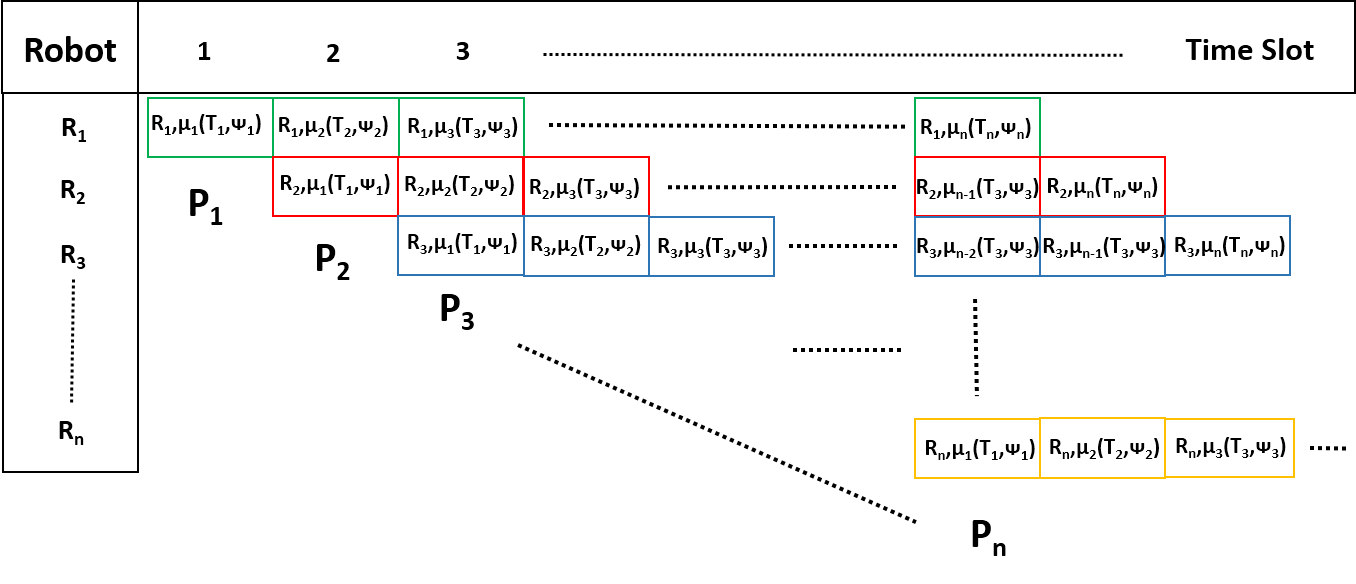}}
	\caption{Pipelined execution of a set of sequential and interdependent tasks having shared resources in the proposed CPS where different colours indicates different robots}
	\label{fig:two}
\end{figure}
\noindent\textbf{Objective 3.}
The final objective of our work is to facilitate the concept of \textit{on-the-fly} ordering. In our proposed CPS scenario, one may require to modify, add (insert) or delete tasks in/from the set \textit{\textbf{T}} \textit{on-the-fly} while the robots are executing tasks in the current set \textit{\textbf{T}}. Further, it may happen that the robots themselves, which are analogous to the processing elements in a pipeline, need to be inserted or removed due to failures, low batteries etc. Hence, providing such flexibilities to the end users of such a system is of vital importance.

Taking all these challenges into consideration, we have converted the graph in Figure~\ref{fig:one} into a pipeline model as depicted in Figure~\ref{fig:two}. The vertical axis in herein represents the robots in action while the horizontal one represents the time slots when the robot $R_{n}$ uses the resource $\psi$$_{n}$ in order to accomplish task $T_{n}$. In addition, $P_1, P_2, P_3,..,P_n$ denotes pipelines formed at time slots 1, 2, 3,...,n respectively. As discussed earlier, it can be seen that in time slot 2, the program carried by $\mu$$_{2}$ facilitates $R_{1}$ to execute $T_{2}$ using the resource $\psi_{2}$. Concurrently, $R_{2}$ executes $T_{1}$ using the program in $\mu$$_{1}$ and the associated resource $\psi_{1}$. Both $\mu_{1}$ and $\mu_{2}$ remain resident on the respective robots $R_{2}$ and $R_{1}$ until the tasks are accomplished and thus are not available to any other robot during slot 2, thus ensuring mutual exclusion among the robots. It may also be noted that $R_{1}$ executes $T_{3}$ whose program is carried by $\mu_{3}$ using $\psi_{3}$ in time slot 3 only after $R_{1}$ and $R_{2}$ both have executed $T_{2}$ and $T_{1}$ respectively as shown in Figure~\ref{fig:one}. It can be seen that in the $n^{th}$ time slot, the pipeline becomes full with all robots $R_{1}$, $R_{2}$,\dots,~$R_{n}$ in states $S_{n}$, $S_{n-1}$, $S_{n-2}$,\dots,~$S_{1}^\star$ executing the allocated tasks $T_{n}$, $T_{n-1}$, $T_{n-2}$,\dots,~$T_{1}$ using the associated resources $\psi$$_{n}$, $\psi$$_{n-1}$, $\psi$$_{n-2}$,\dots,~$\psi$$_{1}$ respectively without any contention. It may be noted that these associated resources could be subsets of $\psi$, namely $\psi^{n}$, $\psi^{n-1}$, $\psi^{n-2}$,\dots,~$\psi^{1}$.

In the next section, we present the mechanism to achieve the objectives listed above.

\section{The Proposed Mechanism}
In the CPS scenario used herein, the robots initially reside at a bay or docking station and their states are initialized to $S_{1}^\star$. This signifies that all the robots in \textbf{\textit{R}} are currently vying to execute the associated task viz. $T_{1}$. As mentioned earlier, every robot hosts an agent framework that allows these agents to knit through them. 

 In addition, all the mobile agents in $\boldsymbol\mu$ are also released into the network \textit{\textbf{W}} by the Job Distributor $J_{Dist}$. Task ordering among the jobs is highly dependent on how the $J_{Dist}$ assigns tasks and their associated resources to the agents. Thus, a separate section is provided for its discussion along with an underlying algorithm. 

\subsection{Job Distributor}
The Job Distributor $J_{Dist}$ assigns the various subsets (\{$T_i,\psi^i$\}) within a job to corresponding agents along with the information on the associated set of resources required to execute the tasks. It also embeds the State Information (SI) of the robots it can serve together with next state to which the robots need to transit. Further, the $J_{Dist}$ also maintains a list of already assigned resources so that the same resource is not assigned to other agents. An agent returns to the $J_{Dist}$ when the assigned task(s) has been executed thereby relinquishing the associated resources.  In the example graph shown in Figure \ref{fig:one}, a simple scenario was chosen where a task requires only a single resource. But in a real world, it is natural to have tasks that require multiple resources which may need to be shared with other tasks. Further, the tasks within a job could be sequential or independent. The task assignment strategies followed by the $J_{Dist}$ for different scenarios are described below:\\ \\
\textit{A. Only  sequential tasks} --
\label{s5.1.1}
Consider a warehouse scenario wherein an item  has to be first fetched from a rack, carried to a packing station and finally packed and shipped to its desired destination. This job comprises a total of 3 sequential tasks - \textit{Fetching} ($T_1$), \textit{Carrying} ($T_2$) and \textit{Packing} ($T_3$), all of which are sequential in nature. For cases when these tasks use different resources, the $J_{Dist}$ assigns a distinct agent for each task together with the associated set of resources. 
Thus  $T_1$ and $\psi^i$, are embedded in $\mu_1$. Likewise,   $T_2$ and $\psi^j$ and $T_3$ and $\psi^k$ are embedded within $\mu_2$ and $\mu_3$. The SI and next SI written onto each of these agents $\mu_1$, $\mu_2$ and $\mu_3$ are -- $S_1^{\star} \rightarrow S_2^{T_2} $, $S_2^{T_2} \rightarrow S_3^{T_3}$ and $S_3^{T_3} \rightarrow S_1^{\star}$ respectively. Since $T_3$ is the last task of the job, the next SI is stored as $S_1^\star$ thereby freeing the robot executing this job. These 3 mobile agents are then released into the  network. \\ 
\\
\textit{B. {A mix of sequential and independent tasks}} --
\label{s5.1.2}
Imagine a job that involves a request for exchange of an item. The set of tasks comprising this job could be - first \textit{Fetch} ($T_1$) the new item,  then \textit{Carry} ($T_2$) it to the packing station and then \textit{Pack} ($T_3$) the same. While these tasks are being executed by one robot in a sequential manner, another robot could concurrently \textit{Stamp} ($T_4$) the item as defective (or some such) and \textit{Place} ($T_5$) it  back to the concerned rack. Since the two sequences ($T_1 \rightarrow T_2  \rightarrow T_3$  and $T_4  \rightarrow T_5$) are independent of each other they do not share any resources. For such kind of jobs, the $J_{Dist}$ sets the SI of the assigned agents $\mu_1$, $\mu_2$ and $\mu_3$ to $S_1^{\star}$, $S_2^{T_2}$ and $S_3^{T_3}$ and that of $\mu_4$ and $\mu_5$ to $S_1^{\star}$ and $S_2^{T_5}$, respectively. This parallel execution of the two sequences ($T_1 \rightarrow T_2  \rightarrow T_3$  and $T_4  \rightarrow T_5$) within a job improves both time and robot utilization.  \\
\\
\begin{algorithm}[tbp]
	\caption{Sequence of steps followed by the Job Distributor $J_{Dist}$}
	\label{alg:one}
	\begin{algorithmic}[1]
		\STATE \textbf{Input}: A Job $J_i$ in the form of a set of tasks and associated resources \COMMENT{Jobs can arrive at the Job Distributer asynchronously}
		\STATE \textbf{Output}: A set of mobile agents with each agent containing a task(s) and its associated resource(s)
		\REPEAT
		\IF{All tasks in $J_i$ are Sequential}
		\IF{Resources required are available}
		\IF{Multiple tasks require same set of resources}
		\STATE Follow steps as described in Section \ref{s5.1.1}.C;
		\ELSE
		\STATE Follow steps as described in Section \ref{s5.1.1}.A;
		\ENDIF
		\ELSE
		\STATE Wait for the mobile agents to return and relinquish the resources back to the $J_{Dist}$;
		\ENDIF
		\ELSIF {Tasks in $J_i$ are a mix of Sequential and Independent}
		\IF{Resources required are available}
		\IF{Multiple tasks require same set of resources}
		\STATE Follow steps as described in Section \ref{s5.1.1}.C;
		\ELSE
		\STATE Follow steps as described in Section \ref{s5.1.1}.B;
		\ENDIF
		\ELSE
		\STATE Wait for the mobile agents to return and relinquish the resources back to the $J_{Dist}$;
		\ENDIF
		\ENDIF
		\UNTIL{Job is present}
	\end{algorithmic}
\end{algorithm}

\textit{C. {Multiple tasks using the same resource(s)}} -- 
In the above two cases, we assumed that the tasks that were sequential did not use a common resource i.e. ($\psi^i \cap \psi^j \cap \psi^l = \emptyset$). It may happen that a job comprises two or more tasks which require the same set of resources. Under such conditions, the task assignment is done purely on the basis of the shared resource needed. For example, if $\psi^i$ is required to execute tasks $T_1$ and $T_2$ while $\psi^j$ requires $T_3$. Thus, $T_1$ and $T_2$ are interdependent while $T_3$ is independent (assuming $\psi^i \cap \psi^j = \emptyset$). Under such scenarios, the $J_{Dist}$ assigns both the interdependent tasks $T_1$ and $T_2$ to a single agent $\mu_i$ while $T_3$ is assigned to another agent $\mu_2$.
The State Information (SI) embedded within $\mu_1$ and $\mu_2$ are given below:\\
$\mu_1$: $S_1^\star \rightarrow S_2^{T_2} \rightarrow S_1^\star$ \\
$\mu_2$: $S_1^\star \rightarrow S_1^\star$
\\
It can be seen from the above SI that $\mu_1$ will find a free robot and make it execute $T_1$ and $T_2$ consecutively before freeing it. $\mu_2$ will find a separate free robot and make it execute $T_3$ concurrently. Since $T_1$ and $T_2$ are now within the same agent, mutual exclusion of the resources shared by these tasks are ensured by the agent itself. It may also happen that $\psi^i \cap \psi^j \neq \emptyset$. This can be easily reduced to the scenario similar to $\psi^i$ i.e. all the three tasks $T_1, T_2$ and $T_3$ will become interdependent and thus, packed into a single agent by the $J_{Dist}$.

 It may be noted that in a set of sequential tasks within a job, say, \{$T_1,T_2,T_3$\}, the resource for a certain task(s) ($T_3$) could be free while those of the others are already assigned to agents of the previous jobs. In such scenarios, the $J_{Dist}$ is forced to wait for the agents to return and relinquish the resource(s). However, if $T_3$ is an independent task, the $J_{Dist}$ will assign it to a separate agent and release it. The agent then follows Algorithm \ref{alg:two} and executes the assigned task. The algorithm for the $J_{Dist}$ is portrayed in Algorithm \ref{alg:one}.

\subsection{Mobile Agent based Mechanism}

Consider a scenario with repetitive jobs with similar tasks and associated resources are landing on the $J_{Dist}$ which are then assigned to the corresponding mobile agents and release into the network of robots. Now, as soon as the mobile agent $\mu_{1}$ lands on robot (say $R_{1}$), it verifies the current state of that robot. If a matching state is found (which in this case is $S_1^\star$), $\mu_{1}$ provides the code for the task $T_1$ to the robot ($R_1$ here). Hence, the robot $R_1$ commences the execution of task $T_{1}$ by acquiring the resources $\psi_{1}$ as per the program received from the agent $\mu_1$. After the execution of task $T_{1}$ by $R_{1}$, the mobile agent $\mu_{1}$ updates the state of $R_{1}$ to the next state (depending upon the next task). Consequently, $R_{1}$ relinquishes the resource $\psi^{1}$ and waits for $\mu_2$ to arrive. The mobile agent $\mu_{1}$ then leaves the robot, returns back to $J_{Dist}$ and releases the task along with the associated resource information. This task and resource is then assigned to a new mobile agent for the next job (which in the current scenario is similar to the previous job) by the $J_{Dist}$  and is then released into the network. If $\mu_1$ does not find a matching state, it migrates to another neighbouring robot in a conscientious manner, thereby avoiding more recently visited robots.

The mobile agent $\mu_{1}$ for job $J_2$ lands up in another robot (say $R_{2}$) in state $S_{1}^\star$ and makes it execute task $T_{1}$ using the resource $\psi^{1}$. In this manner, $\mu_{1}$ continues to make all robots in state $S_{1}^\star$ to perform task $T_{1}$ sequentially. When $\mu_{2}$, which is also migrating within the network, lands in $R_{1}$, it aids the latter in the execution of task $T_{2}$ using $\psi^{2}$. Hence, both the robots $R_{1}$ and $R_{2}$ execute the tasks $T_{2}$ (job $J_1$) and $T_{1}$ (job $J_2$) respectively in a concurrent manner forming a 2-stage pipeline. As time progresses all the \textit{k} robots start executing distinct tasks concurrently to form of a \textit{k}-stage pipeline. Here, the autonomous mobile agents act as tokens to acquire the associated resources in order to carry out an execution. Algorithm~\ref{alg:two} depicts the steps that each mobile agent follows for the execution of their assigned tasks. Thus, it can be seen that by virtue of following this algorithm,  the set of agents $\boldsymbol\mu$ order the execution of tasks, in a manner that ensures mutual exclusion of shared resources among the jobs.

\begin{algorithm}
	\caption{Sequence of steps followed by each mobile agent $\mu_i$ for the execution of their assigned task $T_i$}
	\label{alg:two}
	\begin{algorithmic}[1]
		\STATE \textbf{Input}: State $S_x \in$ \{$S_1^{\star},S_x^{T_i}$\} and Program for task $T_{i}$ $\in$ \textit{\textbf{T}}; \COMMENT{ State is $S_1^\star$ if $T_i$ is the first task to be executed else State is $S_x^{T_i}$}
		\STATE \textbf{Output}: Execution of task $T_i$, $\forall R_k$ $\in$ \textbf{\textit{R}}; 
		\REPEAT
		\STATE migrate\_to($R_k$) ; \COMMENT{Agent migrates to a robot $R_k$}
		\STATE $S$=get\_state($R_k$) ; \COMMENT{Agent fetches the current state of robot $R_k$}
		\IF{$S_x$==$S$}
		\STATE commence\_execution($T_i$,$\psi ^i$) ; \COMMENT{Agent makes robot $R_k$ execute the code for $T_i$ using $\psi_i$}
		\STATE $S_{x+1} = $ get\_next\_state(); \COMMENT{Agent calls the function to get the next State Information (SI) stored within it}
		\STATE update\_state($R_k$, $S_{x+1}$); \COMMENT{Agent updates the state of the robot $R_k$ to the next state carried by the agent}
		\ENDIF
		\STATE leave\_robot($R_k$) ; \COMMENT{Agent migrates into the network to search for other robots} 
		\UNTIL{Job is present}
	\end{algorithmic}
	Note: A mobile agent carries with it the program or code for a task $T_i$ assigned to it, the state $S_x$ of the robots which it needs to search for and execute the code for $T_i$ along with the very next state the robot should transit (after execution of $T_i$), in accordance with the job whose task $T_i$, it carries.
\end{algorithm}

The proposed solution ensures that the free robots are selected and mutually excluded once they start a task within a specific job. Thus, once a robot is booked (mutually excluded) for a job by an agent, the same robot continues to execute all tasks related to this job. The robot is finally released only after the last task (contained within the related agent) is executed. Mutual exclusion is also taken care of when tasks common to multiple jobs require the same resource. Common tasks requiring different resources occurring across multiple jobs are executed concurrently.   
Mobile agents, once released into the network, act autonomously without any central control. With many networked robots in the scenario and with mobile agents knitting through this network,  this proposed CPS as a whole, performs in a decentralized and distributed manner.   
\subsection{Asynchronous Execution Times}

Unlike instructions in a CPU, the tasks executed by robots may not have fixed execution times. This could be due to a range of reasons which include wear and tear of various parts, the nature of the paths traversed by a robot, obstacles, charging times and network delays. This issue of non-uniformity in execution times of the various tasks in the pipeline cannot be efficiently handled by the traditional method of using a common clock. 

In the present decentralized and distributed CPS, a mobile agent is the only entity that has the code for the execution of a specific task. To mitigate the problem of varying time periods in the robotic pipeline, the mobile agents do not leave a robot until the concerned task is accomplished. Consider a case when $R_{i}$ is executing task $T_{j}$ using $\mu_{j}$ and $T_{j}$ takes more time than $T_{j-1}$ which is being executed by $R_{i+1}$ using $\mu_{j-1}$. This forces $R_{i+1}$ (after the execution of $T_{j-1}$) to wait for the completion of execution of $T_{j}$ by $R_{i}$. This is because $\mu_{j}$ (currently within $R_{i}$) has not yet been released. Thus, even though the time durations that the robots take to switch from one task to another within the pipeline keep varying over time, the mobile agents facilitate pipelined execution without the use of a common clock. This makes the proposed mechanism adaptive to varying task execution times. 

\subsection{Addition/Deletion of task(s) \textit{on-the-fly}}
\label{otfp}
A real-world system is always prone to changes which could be sudden or gradual. For a system comprising sequential tasks, these changes can be in the form of addition of new tasks or the deletion of already existing tasks to/from the set \textit{\textbf{T}}. There may also be a case where an existing task is required to be replaced by a new or modified version. Traditionally in a centralized system, one would have to bring the whole system down by suspending the executions of all the tasks and then restart the same after the modifications are made. This naturally is a time-consuming and inefficient exercise. The proposed method for the execution of sequential tasks inherently allows for \textit{On-The-Fly Programming} (OTFP) without bringing the system down. In order to ensure the modification of the task sequence, all the state transitions, from one state to the next, are stored \textit{a priori} locally in the \textit{state transition database} of each robot. In this context, modification could mean addition, deletion or altering the sequence in which the tasks are executed.

The addition of a new task to the set \textit{\textbf{T}} requires two new mobile agents \textemdash\ one that updates the state change information in the robots (referred to as the Sequence Agent ($\mu_{seq}$)) and another that carries the program for the new task ($\mu'_1$). The former agent, $\mu_{seq}$, which is released into the network \textit{\textbf{W}} with the new modified sequence, migrates within \textit{\textbf{W}} and updates the state transition database within each robot to reflect the modifications. Thus, if the initial state transition database in all robots comprised the sequence $S_{1}^\star$, $S_{2}$, $S_{3}$,\dots, $S_{n}$ and the new task to be inserted between $T_{1}$ and $T_{2}$ is $T'_{1}$ then this agent updates the sequence to $S_1^\star,S'_1,S_2,S_3,\dots,S_n$ in all robots in $R$. This would mean that a robot completing the execution of task $T_1$ (in state $S_1^\star$) would now transit to $S'_1$ instead of $S_2$ thereby executing the associated task $T'_1$ before $T_2$ using the second newly released agent $\mu'_1$. Once the modifications are done in each robot, the $\mu_{seq}$ terminates itself. The second agent $\mu'_1$ is the one that carries the new program as its payload and aids the robots to perform the new task $T'_1$. This agent behaves the same way as all the other agents in the set $\mu$. 

Deletion is done by merely deleting the concerned state in the transition database by this agent. It may be noted that if any of the task(s) previous to the task currently being executed by the robot gets modified, then the robot continues with the successive tasks and does not redo the entire job.
The sequence can also be altered in a similar manner to control the order in which the tasks in \textit{\textbf{T}} are executed. Both addition and deletion, thus facilitate the shuffling of the sequence of tasks in the pipeline \textit{on-the-fly}. The above feature thus provides OTFP facility to the system.

\subsection{Mutual Exclusion for Parallel Tasks}
Contrary to the pipelined case, all tasks in the set \textit{\textbf{T}} can be executed concurrently in a fully parallel CPS. Since each task $T_{i}$ has its own dedicated resource $\psi_i$, the agent $\mu_i$ can latch on to any robot $R_{j}$ (provided it is free) and commence executing the associated task. Thus, if there are \textit{n} tasks (i.e. \textit{n} agents) and \textit{n} robots then at any point of time all agents can execute their respective tasks using a robot each. If there are \textit{m} jobs comprising \textit{n} tasks each and if all tasks take the same amount of time \textit{t} for execution, then the total time required for execution of all the jobs would be \textit{m*t}, where * designates the multiplication operator. Mutual exclusion will be preserved, even if the number of robots is greater than the number of jobs. This is so since each task is associated with a single agent which in turn can use only one robot at any moment of time. It may thus be noted that the mechanism described herein can cater to both sequential and parallel sets of tasks.

\subsection{Deadlock Freedom}
According to Coffman et al. \citeyear{Coffman:1971:SD:356586.356588}, a system is in a deadlock state if all the four conditions defined below hold simultaneously --
\begin{enumerate}
    
\item Mutual Exclusion: The resources required are non-shareable and thus requires mutual exclusion.
\item No-Preemption: Resources already assigned cannot be preempted.
\item Circular wait: Presence of circular list or chains of processes waiting for resources acquired by their predecessors.
\item Hold and Wait: A process is holding at least one resource and is also waiting to occupy another resource.
\end{enumerate}

The conditions (1) and (2) hold for the current proposed system. Mutual exclusion is a necessary requirement since the resources become non-shareable once the robots latch on to them. Preemption comes with the risk of indefinite starvation of a resource(s) by the robot preempted by the system and thus adds to the overall cost of execution.

Consider the resource-allocation graph shown in Figure \ref{fig:one} which has been converted to a pipeline representation portrayed in Figure \ref{fig:two}. According to condition (3), if a resource-allocation graph contains at least one cycle, then it can attain a deadlock state. Thus, in order to show that the proposed system is deadlock free, it is sufficient to prove that the graph is acyclic. By applying Kahn \citeyear{kahn1962topological} algorithm for topological sorting  on the  graph shown in Figure \ref{fig:one}, a pipeline representation similar to Figure \ref{fig:two} can be obtained. This proves that the graph is a Directed Acyclic Graph (DAG). Hence, condition (3) does not hold for the proposed system thereby making it deadlock free. Depending upon the type of job, condition (4) could hold for certain scenarios and therefore does not guarantee the deadlock free behaviour of the proposed system. Even though condition (3) is sufficient to ensure the deadlock free behaviour of the system, further investigations to remove condition (4) could be carried out and forms the part of future work of this paper.
\section{Implementation}
In order to validate the efficacy of the proposed mechanism, we chose an automated warehouse as a CPS in order to implement the proposed mechanism. This CPS is used to process shipments after the orders are received at the warehouse. The CPS within the automated warehouse comprises a set of networked robots and smart racks. The robots are required to fetch items from the racks and deliver them to the packaging zones. These chores can be decomposed into several tasks such as follow a path to the selected rack, pick the item, traverse towards the packaging zone and place the item there. This sequence of tasks involves the use of shared resources such as the racks and the paths. Warehouses generally optimize on space which means that the racks are placed close-by thus allowing only one robot to move in between them. This path as also the concerned rack thus form shared resources which can be used by only one robot at any moment of time. This enforces the need to ensure mutual exclusion of resources within the automated warehouse. 

In order to ensure that mutual exclusion is preserved, warehouse management systems have to either constantly monitor and control the movement of robots or the robots themselves have to manage and regulate such exclusions. The former method is more centralized and resource intensive where a single or multiple set of servers constantly monitor and control the robots. Centralized methods have their own drawbacks \cite{Minar2000}. The latter method, wherein the robots themselves as a whole manage such mutual exclusions and executions, forms a decentralized and distributed approach which is what we portray in this work. 

The performance of the proposed mechanism for ordering task execution was validated by emulation followed by experiments using real robots. In order to test the practical viability over large networks, the proposed method was emulated on real networked nodes. Emulation (and not simulation) was carried out to ensure that the experiment is closer to the real environment and captures the real-time issues such as network failure, congestion, packet/data loss, etc. in the system. According to the authors in \cite{1649164}, emulation offers more concrete and reliable results than simulation. \textit{Tartarus} \cite{Semwal2015}, a mobile agent platform was used for emulation of the proposed mechanism for sequential and interdependent task execution. Each instance of \textit{Tartarus} running on a computer acts as a node in the network. For the experiments, a 100-node network was created with sets of nodes running on separate computers connected through a LAN. A separate node acted as the Job Distributor ($J_{Dist}$) which receives the request (job) for the items and converts it in the form of tasks per request. Another additional PC (personal computer) was used to log the status of all the entities and events during the experiments. These logs were used to plot the graphs and analyze the results. It may be noted that these additional nodes ($J_{Dist}$ and PC) did not participate in the proposed mechanism.

\begin{figure}[t]
	\centering
	\centerline{\includegraphics[width=130mm]{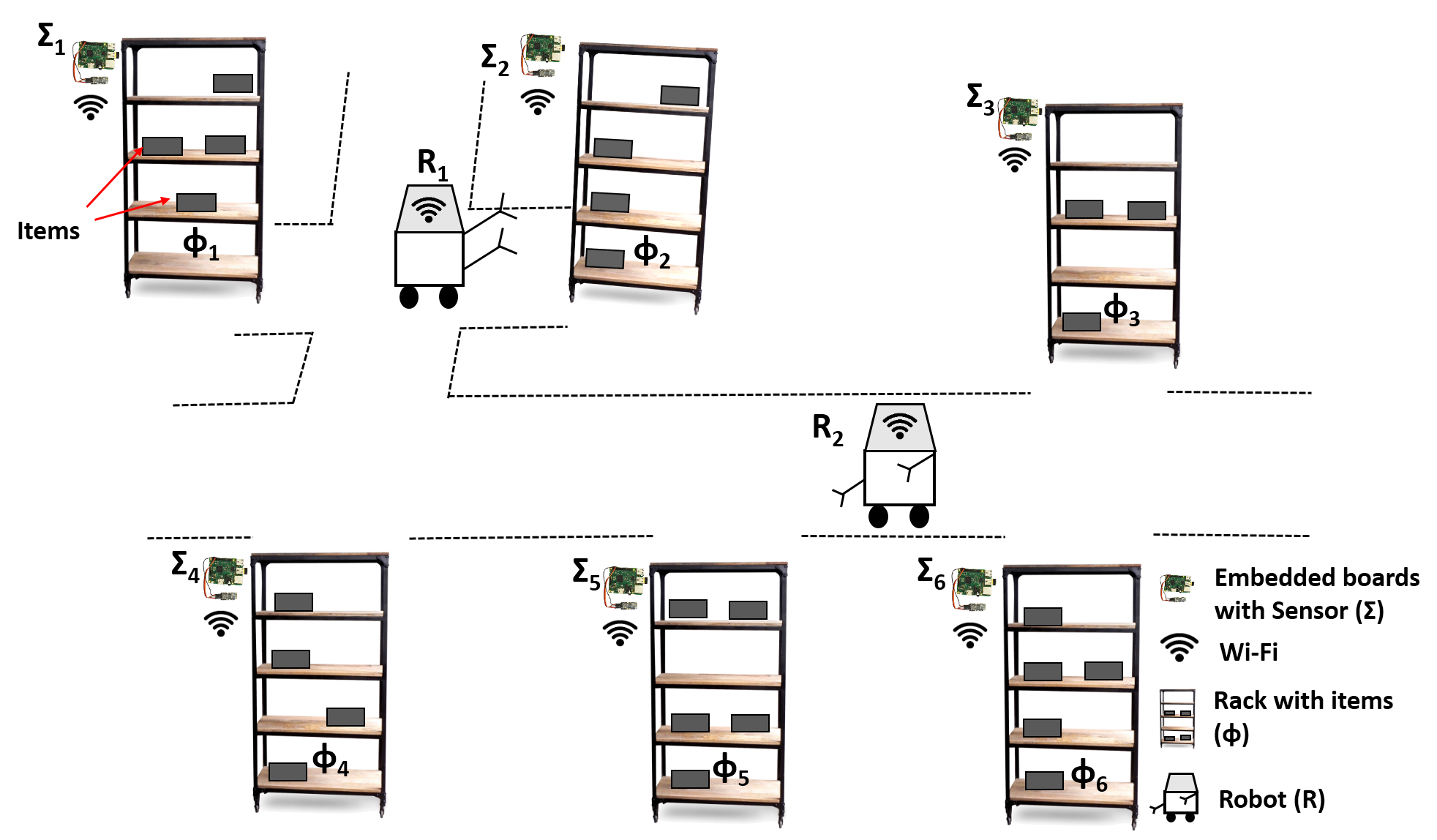}}
	\caption{A warehouse with racks, robots and embedded boards with sensors}
	\label{fig:wms}
\end{figure}
Experiments were conducted to rigorously test the features of the proposed method for scalability, adaptivity and OTFP. Each experiment was performed 10 times and the average of the readings was taken into account while plotting the graphs \footnote{A video of the experiment performed is available at the following link: \\
	\url{http://www.iitg.ernet.in/cse/robotics/?page_id=1993}}.

\section{Experiments and Results}
In this section, we discuss the experiments conducted and results obtained from both emulation and real-robots, separately. The section also compares the proposed approach with its centralized counterpart and highlights the conditions when it is favourable to use the former.

\subsection{Emulation}
\label{sec7.1}

As mentioned earlier, a 100-node \textit{Tartarus} based network was deployed on 10 PCs over a LAN wherein each node represented an instance of \textit{Tartarus}. For honouring equal distribution of load, each PC was initiated with 10 instances of \textit{Tartarus}. Depending on their functionality in the real world, the nodes in the emulation scenario are divided into three types. The \textit{Robotic} nodes (\textit{R}) and the \textit{Shared resource} nodes ($\Phi$) form the \textit{Primary nodes} while the remaining constitute the \textit{Secondary nodes} ($\Sigma$). In the present context, these $\Sigma$ nodes are the inactive nodes which merely allow the agent to flow through in the network such as router, sensor nodes, etc. They may however be made active so as to perform other tasks such as sensing, data processing, etc. based on the application scenario. A static agent residing on each of the R- and $\Phi$-type nodes performed the job of waiting for the mobile agents in order to receive the code for the task to be performed within them. For emulation, the tasks were designed such that it would take $2$ seconds to execute each of them assuming an ideal environment without any unforeseen time lags. Figure~\ref{fig:wms} depicts the primary and secondary nodes for a warehouse scenario.

\begin{figure}[t]
	\centering
	\centerline{\includegraphics[width=110mm]{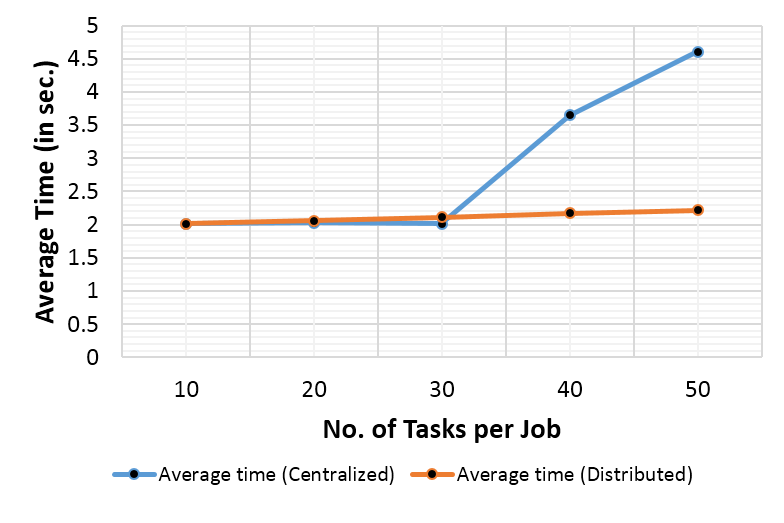}}
	\caption{Centralized versus the proposed decentralized and distributed approach}
	\label{fig:three}
\end{figure}

\subsubsection{Comparison with a Centrally Controlled System }
In order to fortify our stand on the use of this decentralized and distributed mechanism, it is essential to compare the results of the same with those obtained using a centralized control mechanism. A centralized emulation framework for the experimental set-up was thus made using the same agent platform viz. \textit{Tartarus}. A centralized server operating at a node was responsible for posting the relevant commands using TCP-message based communication. This centralized scenario thus comprised the central node hosting the server and the \textit{R} and $\Phi$ nodes acting as its clients. The setup was obviously devoid of mobile agents. The proposed mechanism for execution of mutually-exclusive sequential tasks was emulated on this centralized framework. Instead of the mobile agents carrying the programs required for the execution of the tasks, the robots herein had the required programs to execute all the tasks, embedded in their respective memories \textit{a priori} on-board. The experiment comprised execution of series of jobs with increasing number of tasks. Since the experiment was performed on an emulated framework, a total of 5 different soft computational tasks were chosen viz. Sorting (Sort), Merging (Merge), Addition (Add), Subtraction (Sub) and Division (Div) of data within the nodes. The 5 tasks were repeated for jobs containing more than 5 tasks i.e. if a job has 7 tasks, than the sequential tasks within this job comprises -- Sort, Merge, Add, Sub, Div, Sort, Merge. The tasks were designed in such a way that the total computational time for each task was equal to 2 seconds in an ideal environment without any lags.

Initially, no resources are occupied until the execution commences. The central server thus sends a message to the robot node $R_{1}$ to commence the execution of task $T_1$. In order to ensure mutual exclusion, messages are passed to the central server by each robot as and when a task is to be initiated or completed. On receiving a message from a robot node, say $R_{j}$ after the completion of task $T_{i}$ $(i, j\geq 1$), the centralized server sends a message to the $R_{{j+1}^{th}}$ node informing that the resource $\psi_i$ has been relinquished and that the task $T_{i}$ can now be executed. In this way each robot executes a task $T_{i}$ only when the central server gives it a green signal to do so. The central server thus manages task execution for all robot nodes and hence serves to ensure mutual exclusion.

The variation in the performances of the centralized and the proposed mobile agent based decentralized and distributed mechanisms have been portrayed in Figure~\ref{fig:three}. As can be observed, when the number of tasks per job is below 30, the centralized mechanism seems to perform a tad better than the proposed version. However, as the number of tasks grow (beyond 30), the throughput of the centralized system degrades rapidly since the average time for execution of a task increases. With increasing number of tasks, the volume of information to be exchanged (between the robot nodes and the server) in order to manage the execution of these tasks and ensure mutual exclusion, also increases drastically. Such a large number of server-to-client communications results in a majority of time being wasted on acknowledging and replying to the various nodes. When the central server is loaded with such a large number of requests, it takes more time for the completion of the tasks due to these computational constraints.  In the case of the proposed decentralized approach, the rate of increase in the average time required for completion of the jobs can be observed to be gradual thus indicating the superiority of this approach. These results also show that the system is scalable in the sense that it is hardly affected by the increase in number of tasks per job (aka. mobile agents).

\begin{figure}
	\centering
	\centerline{\includegraphics[width=120mm]{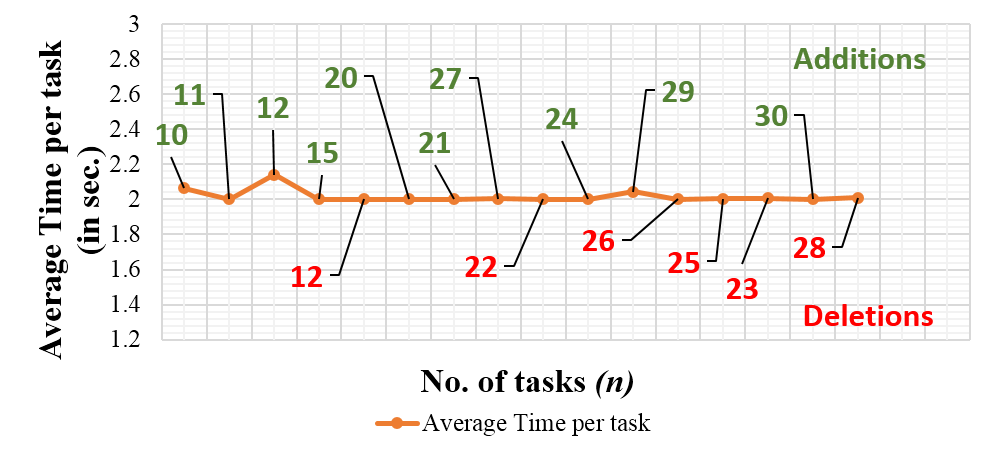}}
	\caption{Addition/Deletion of tasks \textit{on-the-fly}}
	\label{fig:four}
\end{figure}

\subsubsection{Task Addition and Deletion}
As previously mentioned, each of the tasks takes 2 seconds to complete execution in an ideal environment. To study the effect of addition/deletion of tasks \textit{on-the-fly}, tasks were added and deleted during execution. This caused the length of the sequence of tasks in a job to vary at run time. Figure~\ref{fig:four} shows the average time spent per task after addition or deletion of tasks in the set \textit{\textbf{T}}. The numbers above and below the curve in the graph depict the number of tasks comprising the job being executed. As can be seen in Figure~\ref{fig:four}, the experiment initially started off with 10 tasks. While these tasks were being executed, a new task was inserted into the system by the Job Distributor ($J_{Dist}$). This was achieved by following the procedures mentioned in Section \ref{otfp}, eventually modifying the total number of tasks in the task sequence to 11, 12, 15, 12, 20\dots and so on as depicted in Figure~\ref{fig:four}. It may be noted that some of the tasks were also deleted in between the run. One can observe that the graph is almost linear which clearly indicates that, though tasks are added/deleted through OTFP, there seems to be no significant impact on the net time taken for execution. This is so because the extra delays due to communication overheads and computational requirement are distributed among the nodes of the underlying network. This concurrency reduces the effective increase in such lags. In a centralized approach, all these overheads would add up on the controlling entity, thereby degrading its performance.

\subsection{Real-Robot Experiments}
\begin{figure*}[ht]
	\centering
	\subfloat[]{\includegraphics[width = 80mm]{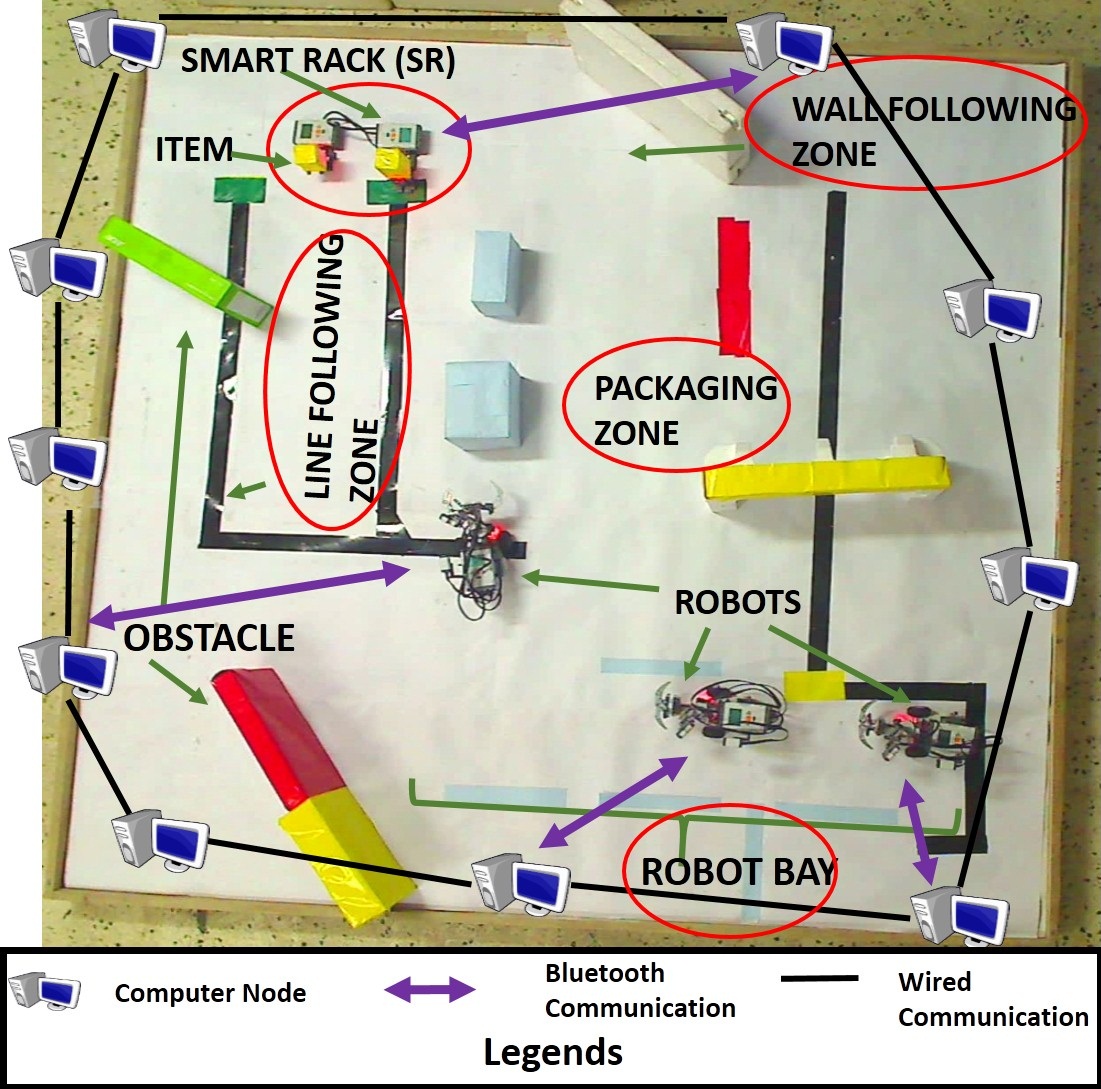} \label{fig:five_a}}
	\subfloat[]{\includegraphics[width = 55mm]{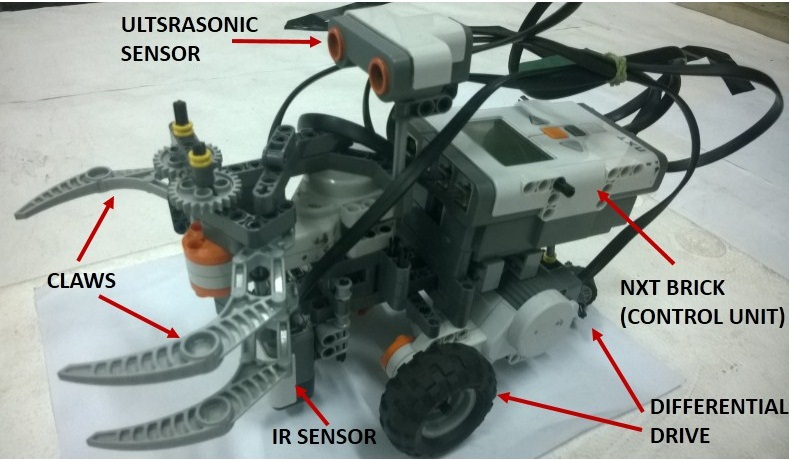} \label{fig:five_b}} 
	\caption{(a) Top view of the test-bed  (b) Structure of one of the LEGO\textsuperscript{\textregistered} MINDSTORMS\textsuperscript{\textregistered} NXT robot used in the experiments
	}
	\label{fig:five}
	
\end{figure*}

In order to provide a proof-of-concept of the actual working of the proposed technique in the real world, a prototype of the warehouse automation scenario was implemented using a set of mobile robots. The job of each mobile robot was to pick an item from a rack, carry it to the packaging zone situated at another location and place it there, from where on it could be parceled and dispatched. As shown in Figure~\ref{fig:five_a}, the experimental setup contains mobile robots and shared resources. The latter include the smart racks and three zones viz. a \textit{line following zone}, a \textit{wall following zone} and a \textit{packaging zone}, all of which need to be shared i.e. they need to be used by the robots in a mutually exclusive manner. A job is divided into four sequential tasks designated $T_1$ to $T_4$, which are briefly described below:

\begin{enumerate}
	\item $T_{1}$: A mobile robot waiting at the robot bay executes $T_1$ by virtue of which it moves forward before it detects a wall. On detecting the wall, it takes a right turn and again moves forward to finally stop when it detects a black line.
	\item $T_{2}$: This task makes the mobile robot to open its claws and start following the black line before a green marker is detected. This marker denotes the location of the smart rack.
	\item $T_{3}$: The robot picks the item from the smart rack using its claws and then follows the wall until a red marker is detected. It then places the item in the packaging zone, thus executing task $T_{3}$.
	\item $T_{4}$: The robot follows a black line before a yellow marker is detected which denotes the start location, thus reentering the bay once again, thereby accomplishing task $T_{4}$.
	
\end{enumerate}

For experimentation with real robots, we used a set of LEGO\textsuperscript{\textregistered} MINDSTORMS\textsuperscript{\textregistered} NXT robots. Since these robots cannot host the \textit{Tartarus} platform within their controller block, they were connected to respective computers (hosting \textit{Tartarus}) via Bluetooth. An interface similar to LPA-PRO-NXT interface \cite{jha2012logic} was used to control the robots via these computers that formed the nodes of the network. The movement of the robots was based on one castor wheel and a two-wheel differential drive. A pair of claws attached to the front of the robots facilitated the picking of items while an on-board ultrasonic sensor gauged the distance of objects in front. The robot was capable of following a black line path using a pair of IR sensors. In addition, a colour sensor detected different markers laid on the floor. Figure~\ref{fig:five_b} shows the structure of one such robot.

As soon as a request for an item is received by the $J_{Dist}$, the same is converted into a job \textit{J} and the corresponding set of mobile agents carrying their respective programs (one per agent) is released into the network. A set of four experiments were performed with the number of robots varying from 1 to 4. A total of 4 jobs were assumed to be always fed into the system. In the first experiment, only one robot $R_1$ was made available at the bay. This signifies a case when one robot needs to perform all the tasks in a sequential manner. \textit{R$_{1}$} which is initially in state $S_{1}^\star$ is thus the only robot available to receive the program for task $T_{1}$ from agent $\mu_1$. After receiving the program, $R_{1}$ starts executing $T_{1}$ since the shared resources viz. all the zones and smart racks are free. As mentioned earlier, a mobile agent resides within the robot before the latter completes the associated task. After the task is completed, the mobile agent leaves the robot and starts migrating into the network in search of another robot in a similar state that requires the associated program.

After task $T_{1}$ is accomplished, $R_1$ goes ahead in the pipeline only if the resources required to execute $T_2$ are available. This is only possible when $R_1$ receives the program to execute $T_2$ via the corresponding agent $\mu_2$. Task $T_{2}$ is that of picking an item from the rack. The third task $T_{3}$ is to place this item at the packaging zone. After $T_{3}$ is accomplished $R_{1}$ performs the final task $T_4$ of following the black line and returning to the bay from where it started. This experiment is one of the conventional ways of performing a sequence of tasks with a single robot. It can thus be used as a baseline while comparing the results obtained from experiments using the proposed decentralized and distributed method using pipelined execution. 

In the second, third and fourth set of experiments, the same jobs were executed but now in a pipelined manner with the numbers of robots enumerated as 2, 3 and 4 respectively.

\begin{figure*}[ht!]
	\centering
	\subfloat[]{\includegraphics[width = 72mm]{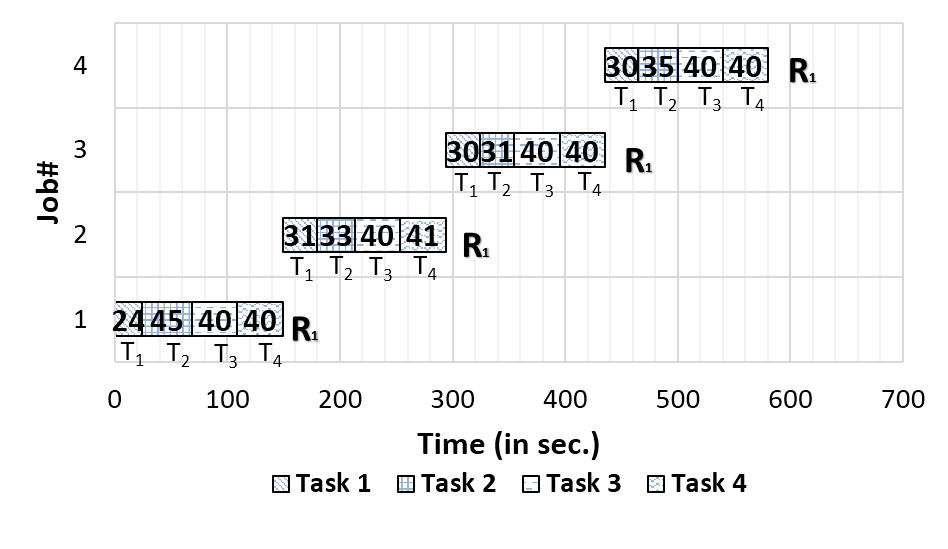} \label{fig:six_a}} 
	\subfloat[]{\includegraphics[width = 72mm]{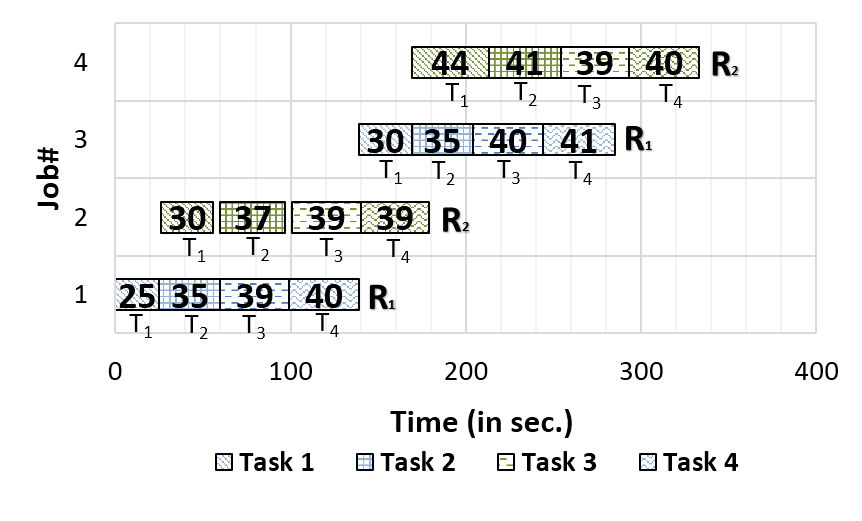} \label{fig:six_b}} 
	\ \subfloat[]{\includegraphics[width = 72mm]{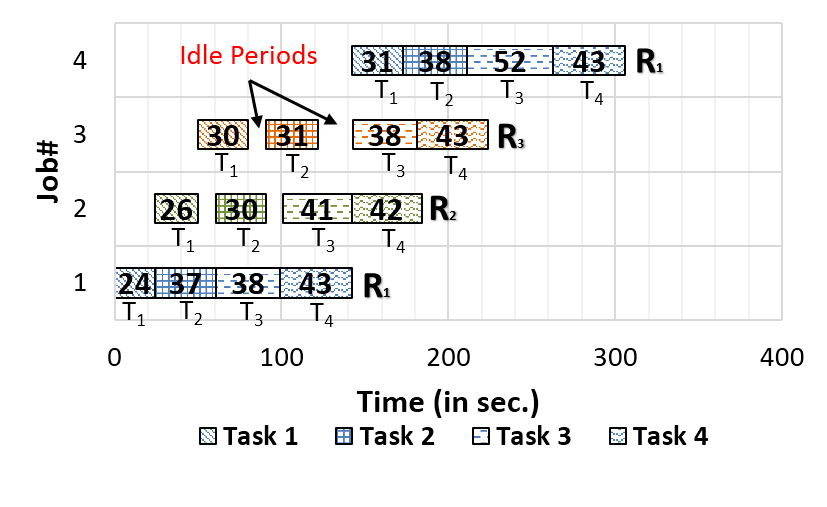} \label{fig:six_c}} 
	\subfloat[]{\includegraphics[width = 72mm]{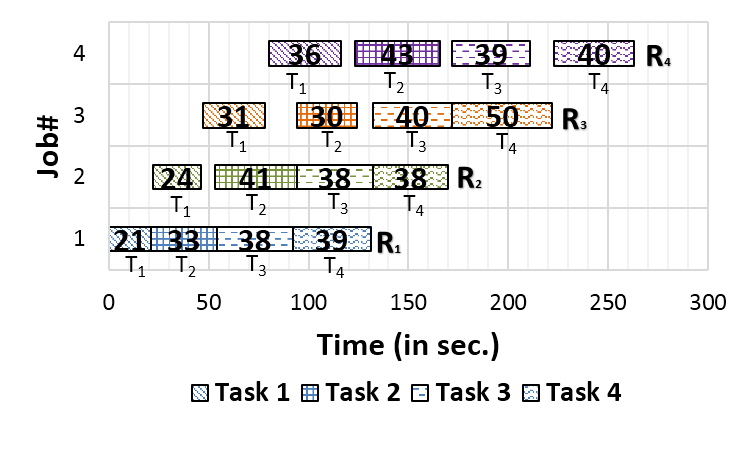} \label{fig:six_d}} 
	\caption{(a) Execution of tasks using a single robot (b), (c) and (d) Pipelined execution for varying number of robots i.e. 2, 3 and 4 respectively}
	\label{fig:six} 
\end{figure*}

Figure~\ref{fig:six} depicts four graphs (viz. \ref{fig:six_a}, \ref{fig:six_b}, \ref{fig:six_c} and \ref{fig:six_d}). Each graph herein corresponds to the results obtained when the number of robots were varied from 1 to 4 respectively. The X-axis in each graph represents the time consumed (in seconds) by the tasks while the Y-axis represents the jobs to be done. The numbers imprinted on the boxes within the graphs denote the execution times taken by the corresponding tasks.

Figure~\ref{fig:six_a} shows the results obtained when there is only one robot available to complete the jobs. The different patterns filled inside the boxes in each row denote the corresponding tasks viz. $T_1$, $T_2$, $T_3$ and $T_4$. As can be seen in the Figure~\ref{fig:six}, each row of boxes i.e. tasks denotes a specific job. As soon as all the tasks comprising a job are completed, the robot switches to executing the first task of the next job. In the case when there is only a single robot there are no inter-job execution delays. This is so because the robot does not have to wait for any shared resources to become free for its use.

Figures~\ref{fig:six_b}, \ref{fig:six_c} and \ref{fig:six_d} show results for cases when the number of robots available are 2, 3 and 4 respectively, in the experiments. In the Figure~\ref{fig:six_b}, two distinct colours have been used to show the two robots \textemdash\ blue signifying $R_1$ while green representing $R_2$. It can be observed that when $R_1$ is executing $T_1, R_2$ waits in the robot bay before the resource $\psi_1$ acquired by $R_1$ is relinquished. When $R_1$ completes the task $T_1$, it relinquishes the associated resource $\psi_1$, which in turn triggers $R_{2}$ to commence execution of $T_{1}$ while $R_{1}$ switches to execute $T_{2}$ using resource $\psi_2$. Execution commences only after the associated agent ($\mu_1$ for $T_1$ and $\mu_2$ for $T_{2}$) reach the concerned robot and provide the relevant programs. Thus, when the mobile agent $\mu_1$ triggers execution of $T_{1}$ by $R_{1}$, the rest of the robots at the bay cannot execute $T_{1}$ since the associated mobile agent $\mu_1$ is now busy with $R_1$. This inherently ensures proper ordering of execution of tasks while also obeying mutual exclusion.

Subsequently, both $R_1$ and $R_2$ enter the pipeline and concurrently execute tasks $T_2$ and $T_1$ respectively. It can be observed that the robots seem to take unequal times to execute the same tasks which in turn cause \textit{idle periods} between the executions of two consecutive tasks. Each \textit{idle period} between the tasks along the row indicates the extra time the robot waits for the resource of the subsequent task to be relinquished by its predecessor. In the real world, execution times depend on the wear and tear that the robots undergo, their controllers, charge on the battery and other environmental conditions. In Figures~\ref{fig:six_c} and \ref{fig:six_d}, these \textit{idle periods} are more prominent due to the presence of more robots. Thus, one cannot provide guarantees that a task will take the same amount of time to complete as it did earlier as in the emulation experiments. The speed-up obtained when 2, 3 and 4 robots was used were found to be 1.75, 1.84 and 2.21 respectively.


\subsubsection{A mix of Sequential and Independent tasks within a job}
\begin{figure}[t]
	\centering
	\centerline{\includegraphics[width=90mm]{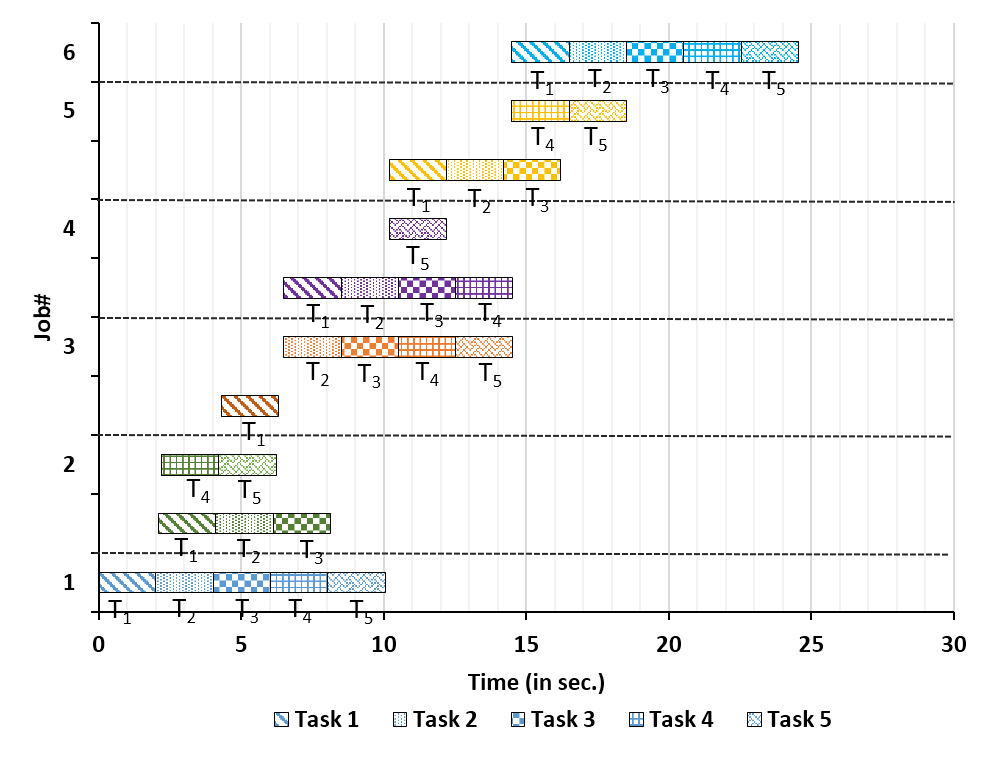}}
	\caption{Execution of jobs comprising both sequential and independent tasks}
	\label{fig:seq_ind}
\end{figure}

Real-world jobs usually comprise heterogeneous tasks where a few of them could be sequential while the rest may be independent. Results for the experiments conducted for such jobs have been portrayed in Figure~\ref{fig:seq_ind}. The experiment was conducted in an emulated environment similar to the one discussed in Section  \ref{sec7.1}. As can be seen from the figure, job $J_1$ and $J_6$ are  composed of purely sequential tasks and thus are executed in sequence. Jobs $J_2$, $J_3$, $J_4$ and $J_5$ comprises two sequences of tasks which are independent of each other. Thus, as shown in the figure, the proposed system manages to execute the two sequences of tasks within a  job in sub-optimal parallel manner.

\subsubsection{Modification of the Sequence of Tasks}
\begin{figure}[t]
	\centering
	\centerline{\includegraphics[width=110mm]{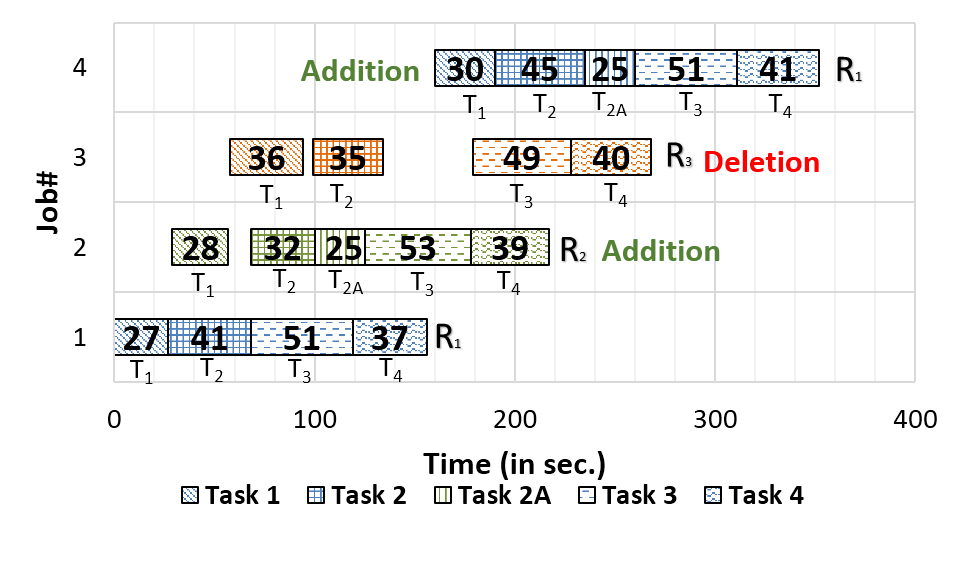}}
	\caption{Execution of tasks using 3 robots - Addition and Deletion of task $T_{2A}$ \textit{on-the-fly}}
	\label{fig:seven}
\end{figure}

Figure~\ref{fig:seven} shows the graph when a task is added and deleted \textit{on-the-fly} from the sequence of tasks $T$. In this experiment, a new task $T_{2A}$ was introduced via $\mu_{2A}$ during the execution of $T_{2}$ in Job 2. It was then deleted immediately after its execution in Job 2. The same task was again added during the execution of $T_{2}$ in Job 4. The additions and deletion were performed using mobile agents as described in Section \ref{otfp}. The new task in the context of our warehouse scenario was a detour from the normally used path which constituted the mobile agent $\mu_{2A}$ and the associated resource $\psi_{2A}$. After $R_{2}$ completed task $T_2, \mu_{2A}$ caused it to execute $T_{2A}$ in Job 2. Since $\psi_2$ is now free, $R_{3}$ (saffron coloured row box) commenced execution of $T_{2}$ using $\mu_2$ concurrently, as seen in Figure~\ref{fig:seven}. It can be seen that the introduction of task $T_{2A}$ introduced a large \textit{idle period} in Job 3. This was because when $R_{3}$ was executing task $T_{2}$, $R_{2}$, having finished execution of $T_{2A}$, commenced the execution of $T_{3}$. The case arose because the time for executing $T_{2A}$ by $R_{2}$ was less than that for executing $T_{2}$ by $R_{3}$. Addition and deletion of tasks \textit{on-the-fly} seemed to have no effect on the other concurrently running tasks.

\subsubsection{Removal of robots}
\begin{figure}[ht!]
	\centering
	\centerline{\includegraphics[width=110mm]{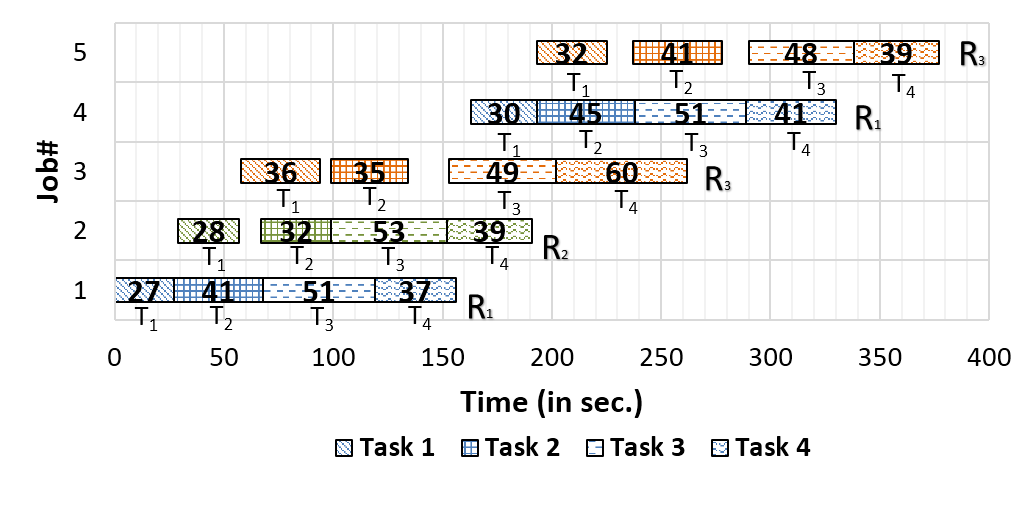}}  
	\caption{Execution of tasks using 3 robots - Removal of a robot ($R_{2}$) \textit{on-the-fly}}
	\label{fig:eight}
\end{figure}

The batteries of all the robots needs to be recharged after their energy levels go below a certain predefined threshold. The threshold was chosen in such a way that even if this threshold were to be crossed at the commencement of a job, the robot would have enough energy to complete all the remaining tasks in that job and then return to the robot bay for charging. This is synonymous to the removal of a robot from the system. It can be observed in Figure~\ref{fig:eight}, that $R_{2}$ was removed after it completed Job 2 in a 3-robot scenario. The absence of $R_{2}$, caused $R_{3}$ to perform the task $T_{1}$ in Job 5 just after $R_{1}$ completed $T_{1}$ in Job 4. As can be observed the pipeline continues to execute tasks concurrently in spite of the absence of $R_{2}$.

It can thus be observed from the results of the experiments, both emulation and real-world, that though the task execution times vary, the system adapts to these changes and maintains the pipeline even in the absence of a clock. Also, the resources associated to the tasks are used in a manner that mutual exclusion is preserved without any direct communication.

\section{Discussions and Conclusions}
This paper portrays a mechanism for ordering the execution of a set of interdependent tasks within a CPS of robots and sensor nodes operating in the real-world under the constraint of mutual exclusion. The CPS, which forms part of an automated warehouse has been used as the target application. This mechanism, however, can be easily ported to other scenarios where shared resources are utilized and mutual exclusion needs to be implemented. It has also been shown that a centralized solution could no doubt be an option to solving the mutual exclusion problem but its performance degrades as we scale the system. The use of mobile agents makes the mechanism described herein decentralized, distributed and scalable and also allows for changes to be made in the tasks as also the set of robots operating within the system, during run times. The mechanism can also cater to more than a single set of sequential tasks executed concurrently thus forming multiple parallel pipelined sequential tasks where the elements of the pipeline can be shared among other pipelines.

The proposed system is limited by the size of the network of nodes. For very large networks (nodes $>$ 1000), there could be considerable delays while searching for a robot to be serviced by a mobile agent. The situation may worsen (delays may increase) if bandwidth for communication is scarcely available. One solution to control such a limitation is to use pheromone-conscientious migration strategy \cite{godfrey2010pheromone}, which allows for a bi-directional search on part of the robotic nodes and the agents across the network.  Tackling the complexity of the tasks whose programs are carried by the mobile agents, opens up future scope for the research. While on one side a box-pushing operation may need just a single robot, complex tasks, such as carrying a long object, may require multiple robots to synchronize their operations. After reserving the required number of robots, the concerned agent within this proposed system could release a new set of agents or clones \cite{semwal2016tartarus} capable of performing the concerned tasks as also synchronizing them using stigmergy based mechanisms as discussed in \cite{Jha2014}.  

The Job Distributor ($J_{Dist}$) is the only central entity in the
proposed system which could be replaced by its distributed version. This means that instead of a single entity that does the task of book-keeping of the allocated resources, each node could act as an individual but scaled down version of the proposed $J_{Dist}$. These mini versions of the $J_{Dist}$ could then communicate with one another to ensure a distributed locking mechanism. This work can also be extended to scenarios where different robots have different abilities such that some subset of robots can execute only certain tasks. The challenge would then be to choose only those robots which are fit for execution of the required tasks. Since physical processes are performed in the real-world, the entities of a typical CPS are bound to encounter unpredictable changes in the environment. Further, these physical processes inherently operate at a different time-scale leading to the creation of an asynchronous environment within the system.

Both these challenges are addressed inherently, to an extent by the mechanism described in this paper. A robot could fail or be removed from the CPS environment leading to unpredictability of the nature of execution of the pipeline. Unpredictable incidents such as these can conventionally cause the stalling of the pipeline. The absence of a common clock for the pipeline aids the mechanism in coping up with this situation. Since the mobile agents manage the execution and mutual exclusion, the waiting times differ based on the availability of the shared resources which in turn aids in re-synchronizing such unforeseen changes in the environment. This also addresses the challenge posed by fluctuations in the execution times taken by the physical processes or tasks performed by the robots. Additionally, since mobile agents carry the related programs for task execution, this mechanism can support heterogeneity in terms of the use of a range of computational entities including robots and sensor nodes. New entities can be plugged into the CPS along with mobile agents carrying the programs for the associated tasks. The use of mobile agents facilitates the OTFP feature which allows such injection or addition and deletion of tasks on-the-fly. This feature can also help in purging deprecated or faulty programs from the system and replacing them with the correct ones during run time. It is also envisaged to augment this mechanism with Self-healing as in \cite{Jha2014} so that a failure within the CPS is quickly noted and the concerned agent is drawn towards the node so that the task(s) can be re-executed, if required.


%

\begin{acks}

The authors would like to thank the Department of Science and Technology, Government of India, for the funding provided under the FIST scheme to set up the Robotics Lab. at the Department of Computer Science and Engineering, Indian Institute of Technology Guwahati, where the entire work reported herein, was carried out.

The first and second authors would also like to acknowledge the Ministry of Human Resource Development, Government of India, and Tata Consultancy Services (TCS), India, for the financial support provided to them during the course of this work.
\end{acks}


%




\bibliographystyle{ACM-Reference-Format-Journals}

\begin{thebibliography}{00}
	
	
	\ifx \showCODEN    \undefined \def \showCODEN     #1{\unskip}     \fi
	\ifx \showDOI      \undefined \def \showDOI       #1{{\tt DOI:}\penalty0{#1}\ }
	\fi
	\ifx \showISBNx    \undefined \def \showISBNx     #1{\unskip}     \fi
	\ifx \showISBNxiii \undefined \def \showISBNxiii  #1{\unskip}     \fi
	\ifx \showISSN     \undefined \def \showISSN      #1{\unskip}     \fi
	\ifx \showLCCN     \undefined \def \showLCCN      #1{\unskip}     \fi
	\ifx \shownote     \undefined \def \shownote      #1{#1}          \fi
	\ifx \showarticletitle \undefined \def \showarticletitle #1{#1}   \fi
	\ifx \showURL      \undefined \def \showURL       #1{#1}          \fi
	
	\bibitem[\protect\citeauthoryear{Anchal, Saini, and Krishna}{Anchal
		et~al\mbox{.}}{2014}]%
	{saini}
	{Anchal}, {P. Saini}, {and} {C. R. Krishna}. 2014.
	\newblock \showarticletitle{{An Efficient Permission-Cum-Cluster Based
			Distributed Mutual Exclusion Algorithm for Mobile Adhoc Networks}}. In {\em Proceedings of the
		2014 International Conference on Parallel, Distributed and Grid Computing
		(PDGC)}. 141--146.
	
	
	
	\bibitem[\protect\citeauthoryear{Attiya, Kogan, and Welch}{Attiya
		et~al\mbox{.}}{2010}]%
	{localmutex}
	{H. Attiya}, {A. Kogan}, {and} {J. L. Welch}. 2010.
	\newblock \showarticletitle{Efficient and Robust Local Mutual Exclusion in
		Mobile Ad Hoc Networks}.
	\newblock {\em IEEE Transactions on Mobile Computing} {9}, 3 (March 2010),
	361--375.
	
	
	
	\bibitem[\protect\citeauthoryear{Baheti and Gill}{Baheti and Gill}{2011}]%
	{Baheti2011a}
	{R. Baheti} {and} {H. Gill}. 2011.
	\newblock \showarticletitle{{Cyber-physical Systems}}.
	\newblock {\em The Impact of Control Technology\/}  {12} (2011), 161----166.
	
	
	\bibitem[\protect\citeauthoryear{Basagni, Conti, Giordano, and
		Stojmenovic}{Basagni et~al\mbox{.}}{2004}]%
	{giordano2002mobile}
	{S. Basagni}, {M. Conti}, {S. Giordano}, {and} {I. Stojmenovic}.
	2004.
	\newblock Mobile ad hoc networking.
	\newblock {\em John Wiley \& Sons}.
	
	
	
	
	\bibitem[\protect\citeauthoryear{Bellifemine, Poggi, and Rimassa}{Bellifemine
		et~al\mbox{.}}{2001}]%
	{Bellifemine2001}
	{F. Bellifemine}, {A. Poggi}, {and} {G. Rimassa}. 2001.
	\newblock \showarticletitle{{JADE: a FIPA2000 compliant agent development
			environment}}. In {\em Proceedings of the International Conference on Autonomous Agents and
		Multiagent Systems}. 216--217.
	
	
	\bibitem[\protect\citeauthoryear{Botelho and Alami}{Botelho and Alami}{1999}]%
	{Botelho1999}
	{S.~C. Botelho} {and} {R. Alami}. 1999.
	\newblock \showarticletitle{M+: A scheme for multi-robot cooperation through
		negotiated task allocation and achievement}. In {\em Proceedings of the IEEE International
		Conference on Robotics and Automation}, Vol.~2. 1234--1239.
	
	
	
	\bibitem[\protect\citeauthoryear{Boukerche, Machado, Juc\'{a}, Sobral, and
		Notare}{Boukerche et~al\mbox{.}}{2007}]%
	{Boukerche2007}
	{A. Boukerche}, {R.~B. Machado}, {K. R. L. Juc\'{a}}, {J. B. M. Sobral}, {and} {M.~S. M. A. Notare}. 2007.
	\newblock \showarticletitle{{An agent based and biological inspired real-time
			intrusion detection and security model for computer network operations}}.
	\newblock {\em Computer Communications\/} {30}, 13 (2007), 2649--2660.
	
	
	\bibitem[\protect\citeauthoryear{Bulgannawar and Vaidya}{Bulgannawar and
		Vaidya}{1995}]%
	{kmutex}
	{S. Bulgannawar} {and} {N. F. Vaidya}. 1995.
	\newblock \showarticletitle{A distributed K-mutual exclusion algorithm}. In
	{\em Proceedings of the 15th
		International Conference on Distributed Computing Systems}. 153--160.
	
	
	
	\bibitem[\protect\citeauthoryear{Chandy and Misra}{Chandy and Misra}{1984}]%
	{Chandy:1984:DPP:1780.1804}
	{K.~M. Chandy} {and} {J. Misra}. 1984.
	\newblock \showarticletitle{The {Drinking Philosophers Problem}}.
	\newblock {\em ACM Transactions on Programming Language and Systems} {6}, 4 (Oct. 1984), 632--646.
	
	
	
	\bibitem[\protect\citeauthoryear{Chen, Gonzalez, and Leung}{Chen
		et~al\mbox{.}}{2007}]%
	{Chen2007}
	{M. Chen}, {S. Gonzalez}, {and} {V.~C.~M. Leung}. 2007.
	\newblock \showarticletitle{{Applications and design issues for mobile agents
			in wireless sensor networks}}.
	\newblock {\em IEEE Wireless Communications\/} {14}, December (2007), 20--26.

	
	\bibitem[\protect\citeauthoryear{Chertov, Fahmy, and Shroff}{Chertov
		et~al\mbox{.}}{2006}]%
	{1649164}
	{R. Chertov}, {S. Fahmy}, {and} {N. B. Shroff}. 2006.
	\newblock \showarticletitle{{Emulation versus Simulation: A case study of
			TCP-targeted denial of service attacks}}. In {\em Proceedings of the 2nd International
		Conference on Testbeds and Research Infrastructures for the Development of
		Networks and Communities, (TRIDENTCOM'06)}. 10 pp.--325.
	
	
	
	\bibitem[\protect\citeauthoryear{Cruz-Cunha}{Cruz-Cunha}{2011}]%
	{cruz2011handbook}
	{M.~M. Cruz-Cunha}. 2011.
	\newblock { {Handbook of Research on Mobility and Computing: Evolving
			Technologies and Ubiquitous Impacts}}. (2 Volumes).
	\newblock {\em IGI Global},1--1584.

	
	
	\bibitem[\protect\citeauthoryear{Dhenakaran and Parvathavarthini}{Dhenakaran
		and Parvathavarthini}{2013}]%
	{Dr.S.S.Dhenakaran2013}
	{S. S. Dhenakaran} {and} {A. Parvathavarthini}. 2013.
	\newblock \showarticletitle{{An Overview of Routing Protocols in Mobile Ad-Hoc
			Network}}.
	\newblock {\em International Journal of Advanced Research in Computer Science
		and Software Engineering\/} {3}, 2 (2013), 251--259.

	
	
	\bibitem[\protect\citeauthoryear{Dias and Stentz}{Dias and Stentz}{2000}]%
	{dias2000free}
	{M.~B. Dias} {and} {A. Stentz}. 2000.
	\newblock \showarticletitle{A free market architecture for distributed control
		of a multirobot system}. In {\em Proceedings of the 6th International Conference on Intelligent
		Autonomous Systems (IAS6'11)}. 115--122.
	
	
	\bibitem[\protect\citeauthoryear{Fife and Gruenwald}{Fife and
		Gruenwald}{2003}]%
	{Fife:2003:RID:776985.776991}
	{L.~D. Fife} {and} {L. Gruenwald}. 2003.
	\newblock \showarticletitle{{Research Issues for Data Communication in Mobile
			Ad-hoc Network Database Systems}}.
	\newblock {\em ACM SIGMOD Record\/} {32}, 2 (June 2003), 42--47.
	
	
	
	\bibitem[\protect\citeauthoryear{Franklin and Graesser}{Franklin and
		Graesser}{1997}]%
	{Franklin}
	{S. Franklin} {and} {A. Graesser}. 1997.
	\newblock \showarticletitle{{Is it an agent, or just a program? A Taxonomy of
			Autonomous Agents}}.
	\newblock {\em Intelligent Agents III\/}  {1193} (1997), 21--36.
	\newblock
	
	
	\bibitem[\protect\citeauthoryear{Gerkey and Mataric}{Gerkey and
		Mataric}{2001}]%
	{Gerkey2001}
	{B.~P. Gerkey} {and} {M.~J. Mataric}. 2001.
	\newblock \showarticletitle{{Principled Communication for Dynamic Multi-Robot
			Task Allocation}}.
	\newblock {\em Lecture Notes in Control and Information Sciences\/}  {271}
	(2001), 353--362.
	\newblock
	
	
	\bibitem[\protect\citeauthoryear{Gerkey and Mataric}{Gerkey and
		Mataric}{2003}]%
	{Gerkey2003}
	{B.~P. Gerkey} {and} {M.~J. Mataric}. 2003.
	\newblock \showarticletitle{{Multi-robot task allocation: Analyzing the
			complexity and optimality of key architectures}}. In {\em Proceedings of the IEEE International
		Conference on Robotics and Automation, ICRA'03}, Vol.~3. 3862--3868.

	
	\bibitem[\protect\citeauthoryear{Godfrey and Nair}{Godfrey and Nair}{2008}]%
	{Godfrey2008}
	{W.~W.~Godfrey} {and} {S.~B. Nair}. 2008.
	\newblock \showarticletitle{{An immune system based multi-robot mobile agent
			network}}.
	\newblock {\em Lecture Notes in Computer Science \/}
	{LNCS} (2008), 424--433.


	
	\bibitem[\protect\citeauthoryear{Godfrey and Nair}{Godfrey and Nair}{2012}]%
	{Godfrey2012}
	{W.~W. Godfrey} {and} {S.~B. Nair}. 2012.
	\newblock \showarticletitle{{A Pheromone based Mobile Agent Migration Strategy
			for Servicing Networked Robots}}. In {\em Proceedings of the 5th International ICST Conference
		on Bio-Inspired Models of Network, Information, and Computing Systems}, 533--541.
	
	
	
	\bibitem[\protect\citeauthoryear{Hadzilacos}{Hadzilacos}{2001}]%
	{groupmutual:2001:NGM:383962.383997}
	{V. Hadzilacos}. 2001.
	\newblock \showarticletitle{A Note on Group Mutual Exclusion}. In {\em
		Proceedings of the Twentieth Annual ACM Symposium on Principles of
		Distributed Computing} {\em (PODC '01)}. 100--106.
	
	
	
	\bibitem[\protect\citeauthoryear{Jha, Godfrey, and Nair}{Jha
		et~al\mbox{.}}{2014}]%
	{Jha2014}
	{S.~S. Jha}, {W.~W. Godfrey}, {and} {S.~B. Nair}. 2014.
	\newblock \showarticletitle{{Stigmergy-Based Synchronization of a Sequence of
			Tasks in a Network of Asynchronous Nodes}}.
	\newblock {\em Cybernetics and Systems\/} {45}, 5 (June 2014), 373--406.

	
	\bibitem[\protect\citeauthoryear{Jha and Nair}{Jha and Nair}{2012}]%
	{jha2012logic}
	{S.~S. Jha} {and} {S.~B. Nair}. 2012.
	\newblock \showarticletitle{{A Logic Programming Interface for Multiple
			Robots}}. In {\em Proceedings of the 3rd IEEE National Conference on Emerging Trends and Applications
		in Computer Science, (NCETACS'12)}. 152--156.
	\newblock
	
	
	\bibitem[\protect\citeauthoryear{Kambayashi, Takimoto, and Kodama}{Kambayashi
		et~al\mbox{.}}{2005}]%
	{Kambayashi}
	{Y. Kambayashi}, {M. Takimoto}, {and} {Y. Kodama}. 2005.
	\newblock \showarticletitle{{Controlling Biped Walking Robots Using Genetic
			Algorithms in Mobile Agent Environment}}. In {\em Proceedings of the  3rd IEEE International
		Conference on Computational Cybernetics, (ICCC'05)}. 29--34.
	\newblock

	
	
	\bibitem[\protect\citeauthoryear{Khaluf and Rammig}{Khaluf and Rammig}{2013}]%
	{Khaluf2013}
	{Y. Khaluf} {and} {F. Rammig}. 2013.
	\newblock \showarticletitle{{Task Allocation Strategy for Time-Constrained
			Tasks in Robots Swarms}}. In {\em Proceedings of the  European Conference on Artificial Life}.
	737--744.

	
	
	\bibitem[\protect\citeauthoryear{Khanna, Singh, and Swaroop}{Khanna
		et~al\mbox{.}}{2015}]%
	{khanna2015dynamic}
	{A. Khanna}, {A.~K. Singh}, {and} {A. Swaroop}. 2015.
	\newblock \showarticletitle{{Dynamic Request Set based Mutual Exclusion
			Algorithm in MANETs}}.
	\newblock {\em International Journal of Wireless and Microwave Technologies
		(IJWMT)\/} {5}, 4 (2015), 1--14.
	\newblock
	
	\bibitem[\protect\citeauthoryear{Suzuki and Kasami}{Suzuki and Kasami}{1985}]%
	{suzuki1985distributed}
	{I.~Suzuki} {and} {T.~Kasami}. 1985.
	\newblock \showarticletitle{A distributed mutual exclusion algorithm}.
	\newblock {\em ACM Transactions on Computer Systems (TOCS)\/} {3}, 4 (1985),
	344--349.
	\newblock
	
	
	\bibitem[\protect\citeauthoryear{Lange}{Lange}{1998}]%
	{Lange1998}
	{D.~B. Lange}. 1998.
	\newblock { Mobile objects and {M}obile agents: The future of distributed
		computing?}
	In \newblock { \em Proceedings of the European Conference on Object-Oriented Programming (ECOOP'98), Lecture Notes in Computer Science}, vol 1445., 1--12.
	
	
	
	\bibitem[\protect\citeauthoryear{Maes, Guttman, and Moukas}{Maes
		et~al\mbox{.}}{1999}]%
	{Maes1999}
	{P. Maes}, {R.~H. Guttman}, {and} {A.~G. Moukas}. 1999.
	\newblock {\em Agents That Buy and Sell}.
	\newblock {\it ACM Communication} {42}, 3 (March 1999), 81--91.
	
	
	
	\bibitem[\protect\citeauthoryear{Matani and Nair}{Matani and Nair}{2011}]%
	{Matani2011}
	{J. Matani} {and} {S.~B. Nair}. 2011.
	\newblock \showarticletitle{Typhon: A {Mobile Agents Framework for Real World
			Emulation in Prolog}}. In {\em Proceedings of the 5th International
		Conference on Multi-Disciplinary Trends in Artificial Intelligence} {\em
		(MIWAI'11)}. 261--273.
	
	
	\bibitem[\protect\citeauthoryear{Coffman, Elphick, and Shoshani}{Coffman
		et~al\mbox{.}}{1971}]%
	{Coffman:1971:SD:356586.356588}
	{E.~G. Coffman}, {M. Elphick}, {and} {A. Shoshani}. 1971.
	\newblock \showarticletitle{System Deadlocks}.
	\newblock {\em ACM Comput. Surv.\/} {3}, 2 (June 1971), 67--78.
	
	\bibitem[\protect\citeauthoryear{Kahn}{Kahn}{1962}]%
	{kahn1962topological}
	{A.~B. Kahn}. 1962.
	\newblock {Topological sorting of large networks}.
	\newblock {\em Communications of the ACM} {5}, 11 (1962), 558--562.
	
		
	
	\bibitem[\protect\citeauthoryear{Minar, Gray, Roup, Krikorian, and Maes}{Minar
		et~al\mbox{.}}{2000}]%
	{Minar2000}
	{N. Minar}, {M. Gray}, {O. Roup}, {R. Krikorian}, {and} {P.	Maes}. 2000.
	\newblock \showarticletitle{{Hive: Distributed agents for networking things}}.
	\newblock {\em IEEE Concurrency\/}  {8} (2000), 24--33.

	
	
	\bibitem[\protect\citeauthoryear{Mostinckx, {Van Cutsem}, Timbermont, Boix,
		Tanter, and {De Meuter}}{Mostinckx et~al\mbox{.}}{2009}]%
	{Mostinckx2009}
	{S. Mostinckx}, {T. {V. Cutsem}}, {S. Timbermont}, {E.~G. Boix}, {\'{E.} Tanter}, {and} {W. {D. Meuter}}. 2009.
	\newblock \showarticletitle{{Mobile-C: A mobile agent platform for mobile C/C++
			agents}}.
	\newblock {\em Software - Practice and Experience\/}  {39} (2009), 661--699.
	
	
	
	\bibitem[\protect\citeauthoryear{Null and Lobur}{Null and Lobur}{2014}]%
	{null2014essentials}
	{L. Null} {and} {J. Lobur}. 2014.
	\newblock { {The essentials of Computer Organization and Architecture}}.
	\newblock {\em Jones \& Bartlett Publishers}.
	
	
	
	\bibitem[\protect\citeauthoryear{Outtagarts}{Outtagarts}{2009}]%
	{outtagarts2009mobile}
	{A. Outtagarts}. 2009.
	\newblock \showarticletitle{Mobile agent-based applications: A survey}.
	\newblock {\em International Journal of Computer Science and Network
		Security\/} {9}, 11 (2009), 331--339.
	\newblock
	
	
	\bibitem[\protect\citeauthoryear{Parker}{Parker}{1998}]%
	{Parker1998}
	{L.~E. Parker}. 1998.
	\newblock \showarticletitle{ALLIANCE: An architecture for fault tolerant
		multirobot cooperation}.
	\newblock {\em IEEE Transactions on Robotics and Automation \/} {14}, 2 (1998),
	220--240.
	\newblock
	
	
	\bibitem[\protect\citeauthoryear{Posadas, Poza, Sim\'{o}, Benet, and
		Blanes}{Posadas et~al\mbox{.}}{2008}]%
	{Posadas2008}
	{J.~L. Posadas}, {J. L. Poza}, {J. E. Sim\'{o}}, {G. Benet}, {and} {F. Blanes}.
	2008.
	\newblock \showarticletitle{{Agent-based distributed architecture for mobile
			robot control}}.
	\newblock {\em Engineering Applications of Artificial Intelligence\/} {21}, 6
	(Sept. 2008), 805--823.
	
	
	\bibitem[\protect\citeauthoryear{Godfrey and Nair}{Godfrey and Nair}{2010}]%
	{godfrey2010pheromone}
	{W.~W.~Godfrey} {and} {S.~B. Nair}. 2010.
	\newblock \showarticletitle{A pheromone based mobile agent migration strategy
		for servicing networked robots}. In {\em International Conference on
		Bio-Inspired Models of Network, Information, and Computing Systems}.
	Springer, 533--541.
	\newblock
	
	\bibitem[\protect\citeauthoryear{Ratnieks and Anderson}{Ratnieks and
		Anderson}{1999}]%
	{Ratnieks1999}
	{F.~L.~W. Ratnieks} {and} {C. Anderson}. 1999.
	\newblock \showarticletitle{Task partitioning in insect societies}.
	\newblock {\em Insectes Sociaux\/} {46}, 2 (1999), 95--108.
	
	
	
	\bibitem[\protect\citeauthoryear{Raynal}{Raynal}{1986}]%
	{raynal1986algorithms}
	{M. Raynal}. 1986.
	\newblock \showarticletitle{Algorithms for mutual exclusion}.
	\newblock {\em The MIT Press, Cambridge, MA\/} (1986).
	\newblock
	
	\bibitem[\protect\citeauthoryear{Semwal, Nikhil, Jha, and Nair}{Semwal
		et~al\mbox{.}}{2016}]%
	{semwal2016tartarus}
	{T. Semwal}, {Nikhil S}, {S.~S. Jha}, {and} {S.~B. Nair}.
	2016.
	\newblock \showarticletitle{TARTARUS: A Multi-Agent Platform for Bridging the
		Gap between Cyber and Physical Systems}. In {\em Proceedings of the 2016
		International Conference on Autonomous Agents \& Multiagent Systems (AAMAS'16)}, 1493--1495.
	\newblock
	
	
	\bibitem[\protect\citeauthoryear{Schumacher}{Schumacher}{2001}]%
	{schumacher2001multi}
	{M. Schumacher}. 2001.
	\newblock \showarticletitle{Multi-agent systems}.
	\newblock {\em Objective Coordination in Multi-Agent System Engineering: Design
		and Implementation\/} (2001), 9--32.
	\newblock

	\bibitem[\protect\citeauthoryear{Semwal, Bode, Singh, Jha, and Nair}{Semwal
		et~al\mbox{.}}{2015}]%
	{Semwal2015}
	{T. Semwal}, {M. Bode}, {V. Singh}, {S.~S. Jha}, {and}
	{S.~B. Nair}. 2015.
	\newblock \showarticletitle{Tartarus: A {Multi-agent Platform for Integrating
			Cyber-Physical Systems and Robots}}. In {\em Proceedings of the 2015
		Conference on Advances In Robotics} {\em (AIR'15)}. Article 20, 6 pages.
	
	
	
	\bibitem[\protect\citeauthoryear{Shi, Wan, Yan, and Suo}{Shi
		et~al\mbox{.}}{2011}]%
	{Shi2011}
	{J. Shi}, {J. Wan}, {H. Yan}, {and} {H. Suo}. 2011.
	\newblock \showarticletitle{{A survey of Cyber-Physical Systems}}. In {\em Proceedings of the 2011
		International Conference on Wireless Communications and Signal Processing (WCSP'11)}. 1--6.

	
	
	\bibitem[\protect\citeauthoryear{Tarau}{Tarau}{1999}]%
	{tarau1999jinni}
	{P. Tarau}. 1999.
	\newblock \showarticletitle{{Jinni: Intelligent mobile agent programming at the
			intersection of Java and Prolog}}. In {\em Proceedings of the 4th
		International Conference on The Practical Application of Intelligent Agents
		and Multi-Agents, (PAAM'99)}, Vol.~99. 109--123.
	\newblock
	
	
	\bibitem[\protect\citeauthoryear{Wu, Zhang, Luo, and Cao}{Wu
		et~al\mbox{.}}{2015}]%
	{wuweigang}
	{W. Wu}, {J. Zhang}, {A. Luo}, {and} {J. Cao}. 2015.
	\newblock \showarticletitle{Distributed Mutual Exclusion Algorithms for
		Intersection Traffic Control}.
	\newblock {\em IEEE Transactions on Parallel and Distributed Systems\/} {26},
	1 (Jan 2015), 65--74.

	
	
	\bibitem[\protect\citeauthoryear{Zaiane}{Zaiane}{2002}]%
	{Zaiane2002}
	{O. R. Zaiane}. 2002.
	\newblock \showarticletitle{{Building a recommender agent for e-learning
			systems}}. In {\em Proceedings of the International Conference on Computers in Education}. 55 -- 59.

\bibitem[\protect\citeauthoryear{Gaber and Bakhouya}{Gaber and
	Bakhouya}{2008}]%
{Gaber2008}
{J. Gaber} {and} {M. Bakhouya}. 2008.
\newblock \showarticletitle{{Mobile Agent-Based Approach for Resource Discovery
		in Peer-to-Peer Networks}}. {\em Agents and Peer-to-Peer Computing}.
63--73.
\newblock

\bibitem[\protect\citeauthoryear{Minar, Kramer, and Maes}{Minar
	et~al\mbox{.}}{1999}]%
{minar1999}
{N. Minar}, {K.~H.~Kramer}, {and} {P.~Maes}. 1999.
\newblock {\em Cooperating Mobile Agents for Dynamic Network Routing}.
\newblock 287--304.


\bibitem[\protect\citeauthoryear{Godfrey, Jha, and Nair}{Godfrey
	et~al\mbox{.}}{2013}]%
{godfrey2013mobileAgntIoT}
{W.~W. Godfrey}, {S.~S. Jha}, {and} {S.~B. Nair}. 2013.
\newblock \showarticletitle{On a Mobile Agent Framework for an Internet of
	Things}. In {\em Proceedings of the 2013 International Conference on Communication Systems and
	Network Technologies}. 345--350.
\newblock

\bibitem[\protect\citeauthoryear{Sempe and Drogoul}{Sempe and Drogoul}{2003}]%
{sempe_evap}
{F. Sempe} {and} {A. Drogoul}. 2003.
\newblock \showarticletitle{Adaptive patrol for a group of robots}. In {\em Proceedings of the 2003 IEEE/RSJ International Conference on Intelligent Robots and
	Systems (IROS 2003) (Cat. No.03CH37453)}, Vol.~3. 2865--2869 vol.3.
\newblock

\bibitem[\protect\citeauthoryear{Chu, Glad, Simonin, Sempe, Drogoul, and
	Charpillet}{Chu et~al\mbox{.}}{2007}]%
{chu_CLInG}
{H.~N. Chu}, {A. Glad}, {O. Simonin}, {F. Sempe}, {A. Drogoul}, {and} {F.
	Charpillet}. 2007.
\newblock \showarticletitle{Swarm Approaches for the Patrolling Problem,
	Information Propagation vs. Pheromone Evaporation}. In {\em Proceedings of the 19th IEEE
	International Conference on Tools with Artificial Intelligence(ICTAI'07)},
Vol.~1, 442--449.
\newblock

	
	
\end{thebibliography}
\end{document}